\documentclass[12pt]{iopart}
\expandafter\let\csname equation*\endcsname\relax
\expandafter\let\csname endequation*\endcsname\relax
\usepackage{amsmath,mathtools}

\usepackage{cite}

\usepackage{graphicx}        % standard LaTeX graphics tool
\usepackage{wrapfig}
\usepackage[bottom]{footmisc}% places footnotes at page bottom

\usepackage{xcolor}

\usepackage{amssymb}

\usepackage{bm,cancel}
\usepackage{accents}
\usepackage{lineno}
\usepackage[T1]{fontenc}

\definecolor{redcol}{rgb}{1.,0.,0.0} 
\definecolor{lnkcol}{rgb}{0.,0.,0.0} %black
\usepackage[pdftex,pdfpagelabels=true,bookmarksdepth=3]{hyperref}
\hypersetup{  colorlinks=true,
              citecolor=lnkcol,%
              linkcolor=lnkcol,%
              urlcolor=lnkcol,
            bookmarksnumbered=true,
            pdftitle = {Computing MHD equilibria of stellarators with a flexible coordinate frame},
            pdfsubject = {},
            pdfauthor = {F.J.Hindenlang, G.G.Plunk, O.Maj},
            pdfkeywords = {}
 }
\usepackage[all]{hypcap}  % hyperlink soll oberhalb des Bilds anhalten
\usepackage[dvips]{thumbpdf}

\definecolor{myblue}{rgb}{0.5,0.,0.5}
\newcommand\hladdedrev[1]{#1} % do not highlight!

\newcommand\hlchanged[1]{#1} % do not highlight!
\newcommand\hlchangedrev[1]{#1} % do not highlight!

\newcommand\dd[2]{\frac{d #1}{d #2}}
\newcommand\ddp[2]{\frac{\partial #1}{\partial #2}}

\newcommand\el{\ell}
\newcommand\elp{\el'}

\newcommand\X{{\bm{X}}_0}
\newcommand\Xp{\X'}
\newcommand\Xpp{\X''}
\newcommand\Xppp{\X'''}
\newcommand\xcont{{{\bm x}|_\zeta}}
\newcommand\Tfrenet{{\bm T}^F}
\newcommand\Nfrenet{{\bm N}^F}
\newcommand\Bfrenet{{\bm B}^F}
\newcommand\TfrenetS{{\bm T}^S}
\newcommand\NfrenetS{{\bm N}^S}
\newcommand\BfrenetS{{\bm B}^S}
\newcommand\Nnew{{\bm N}}
\newcommand\Bnew{{\bm B}}
\newcommand\K{{\bm K}}
\newcommand\hmap{{h}}
\newcommand\epsp{\varepsilon^{+}}
\newcommand\epsm{\varepsilon^{-}}

\newcommand\GFF{$G$-frame}
\newcommand\RZF{$RZ$-frame}
\newcommand\nfp{n_\textrm{FP}}

\newcommand\ttilde{{\tilde{\bm T}_q}}

\newcommand{\thet}{\vartheta}

\newcommand{\Jh}{\mathcal{J}_h}
\newcommand{\Jac}{\sqrt{g}}

\newcommand\linknum[1]{\hlchangedrev{self-linking number{#1}}}

\newcommand\NBplane[1]{$\Nnew,\Bnew$-plane{#1}}
\newcommand\RZplanes[1]{$R,Z$-planes{#1}}
\newcommand\NBplanes[1]{$\Nnew,\Bnew$-planes{#1}}
\newcommand\binormal[1]{bi-normal{#1}}

\begin{document}

\title{\hlchangedrev{Computing MHD equilibria of stellarators with a flexible coordinate frame}}
%OLD
%\title{A \hlchangedrev{general coordinate} frame for computing MHD equilibria in stellarators}

\author{Florian Hindenlang$^{1}$, Gabriel G. Plunk$^{2}$, Omar Maj$^{1}$}
%Stellarator Theory, 
\address{ $^1$Numerical Methods in Plasma Physics, Max Planck Institute for Plasma Physics, Garching, Germany}
\address{$^2$Max Planck Institute for Plasma Physics, Greifswald, Germany}

\ead{florian.hindenlang@ipp.mpg.de}

\begin{abstract} 
For the representation of axi-symmetric plasma configurations (tokamaks), it is natural to use cylindrical coordinates $(R,Z,\phi)$, where $\phi$ is an independent coordinate. The same cylindrical coordinates have also been widely used for representing 3D MHD equilibria of non-axisymmetric configurations (stellarators), with cross-sections, defined in \RZplanes{}, that vary over $\phi$. 

Stellarator equilibria have been found, however, for which cylindrical coordinates are not at all a natural choice, for instance certain stellarators obtained using the near-axis expansion (NAE) \cite{landreman-sengupta-plunk,plunk_landreman_helander_2019}, defined by a magnetic axis curve and its Frenet frame.

\hlchangedrev{In this contribution, we propose an alternative approach for representing the boundary in a fixed-boundary 3D MHD equilibrium solver, moving away from cylindrical coordinates. 
Instead, we use planar cross-sections whose orientation is determined by a general coordinate frame (\GFF). This frame is similar to the conventional Frenet frame, but more flexible. 
As an additional part of the boundary representation, it becomes an input to the equilibrium solve, along with the geometry of the cross-sections.}
We see two advantages: 1) the capability to easily represent configurations where the magnetic axis is highly non-planar or even knotted. 2) a reduction in the degrees of freedom needed for the \hlchangedrev{boundary surface, and thus the equilibrium solver,} enabling progress in optimization of these configurations.

\hlchangedrev{We discuss the properties of the \GFF, starting from the conventional Frenet frame. Then we show two exemplary ways of constructing it, first from a NAE solution and also from a given boundary surface. We present the details of the implementation of the new frame in the 3D MHD equilibrium solver GVEC.} Furthermore, we demonstrate for a highly shaped QI-optimized stellarator that far fewer degrees of freedom are necessary to find a high quality equilibrium solution, compared to the solution computed in cylindrical coordinates.
\end{abstract}

\section{Introduction}
\label{sec:intro}
\hladdedrev{The computation of 3D MHD equilibria of stellarators is a challenging problem. We focus on the so-called fixed-boundary equilibrium solver, that starts from a given closed boundary surface, which acts as a closed flux surface of the plasma region. Thus, the geometrical representation of the boundary surface becomes a crucial part of the equilibrium solve. The main objective of this work is to demonstrate an approach to solving the equilibrium with a generalized and flexible surface representation.}
\hlchangedrev{For this purpose, we introduce curvilinear coordinates that are constructed from two linearly independent vectors attached to a guiding curve. The tangent of the curve and the two vectors form a frame which we denote '\GFF' (general coordinate frame).} 
We claim that the \GFF{} allows us to represent the geometry of stellarators with closed 3D surfaces of basically any shape. 
%
%Similar to the conventional Frenet frame, the \GFF{} smoothly follows a curve in 3D space, with the first basis vector being the tangent unit vector to the curve and the two remaining basis vectors spanning a plane normal to the curve. 
An important requirement for the \GFF{} is that all its basis vectors are \hlchangedrev{\emph{differentiable functions}} everywhere along the curve parameter. The requirement of differentiability is not always met by the conventional Frenet frame, as described in section~\ref{sec:gff}. 

\hladdedrev{The construction of the \GFF{} is flexible and can be adapted to different boundary surface shapes. We provide two examples of its explicit construction. In the first example, we construct it from the Frenet frame of a NAE solution. In the second example, it is constructed from a boundary surface, taken from the publicly available QUASR database \cite{Giuliani_2024_quasr}.}
Another construction method that naturally meets the requirements of the \GFF{} is the Bishop frame \cite{bishop_frame,yilmaz_new_bishop}, which is also called the parallel transport frame, or rotation-minimizing frame, and which can be easily \hlchangedrev{constructed discretely~\cite{rotation_minimizing_frame} from point positions along a 3D curve.}

The main motivation of this work is to use the \GFF{} in the context of the 3D MHD equilibrium solution in stellarators. The conventional and widely used cylinder coordinates, with cross-sections in $R,Z$ parametrized over the geometrical toroidal angle $\phi$, has strong limitations on the 3D surface shapes it can represent, and thus artificially restricts the space of possible shapes that can be explored in stellarator optimization.

One of the most prominent 3D MHD equilibrium codes, founding modern computational stellarator optimization, is VMEC \cite{hirshman_VMEC_1983,Hirshman1986_vmec_free,Hirshman1991_vmec_precond}. A key idea of VMEC is the assumption of closed nested flux surfaces to find 3D MHD equilibria. 
Note that the assumption of a continuum of nested flux surfaces does not necessarily hold in three-dimensional magnetic fields and can lead to current singularities at rational surfaces, as discussed in \cite{Huang_2023}. 
In VMEC, starting from an initial state of closed nested 3D surfaces, expressed in flux-aligned coordinates (or magnetic coordinates) \cite{Dhaeseleer1991}, the equilibrium solution is found by minimization of the MHD energy iteratively, changing the 3D surface shapes while keeping them nested. 
More recently, 3D MHD equilibrium solvers following the same assumption of nested 3D surfaces have been developed, such as DESC\cite{dudt_DESC_2020} and GVEC\cite{gvec-2019}, being different from each other, as well as from VMEC, in their discretization and the way the iterative solution is found. 

When designing GVEC, care was taken separating the actual geometry representation from its toroidal topology. In GVEC, the geometry is described by 2D cross-sections in two coordinates $(q^1,q^2)$ and parametrized by a third periodic coordinate $\zeta$. The mapping from $(q^1,q^2,\zeta)$ to Cartesian coordinates is not fixed and can be described by any expression. Simple examples are the periodic cylinder ($x=q^1,y=q^2,z=\zeta$) and the conventional toroidal representation ($x=q^1\cos(\zeta), y=q^1\sin(\zeta), z=q^2$). It is this feature that allows us to use the \GFF{} in GVEC to solve 3D MHD equilibria.

Stellarator equilibria are traditionally found by numerical optimization of their boundary shapes, to ensure that certain physical properties are satisfied, for example related to plasma confinement and stability.  An alternative and elegant approach to finding optimal stellarator shapes is by use of the near axis expansion (NAE) in Boozer coordinates, as introduced by \cite{garren-boozer-1, garren-boozer-2} and extended over the last years \cite{landreman-sengupta, landreman-sengupta-plunk, plunk_landreman_helander_2019}. Again, nested 3D flux surfaces are assumed, constrained to local MHD equilibrium conditions, and are represented as Taylor expansion around a closed parametrized 3D curve $X(\zeta)$, which is identified with the magnetic axis. Each cross-section is defined over a plane perpendicular to the axis, spanned by normal ($\Nfrenet$) and \binormal{ } ($\Bfrenet$) vectors of the Frenet frame of the magnetic axis. As the configurations are defined by the magnetic axis, generally an arbitrary closed curve, completely new optimized stellarator shapes can be found for which cylindrical coordinates are not at all a natural choice. For instance, certain quasi-symmetric and quasi-isodynamic stellarators, even knotted configurations, have been obtained (or `constructed') using NAE \cite{Landreman_Sengupta_2019, camacho-mata_plunk_2022, plunk2024-QI}.

\begin{figure}[htbp!]
    \centering
    \includegraphics[trim=0 0 0 0,clip,width=0.75\textwidth]{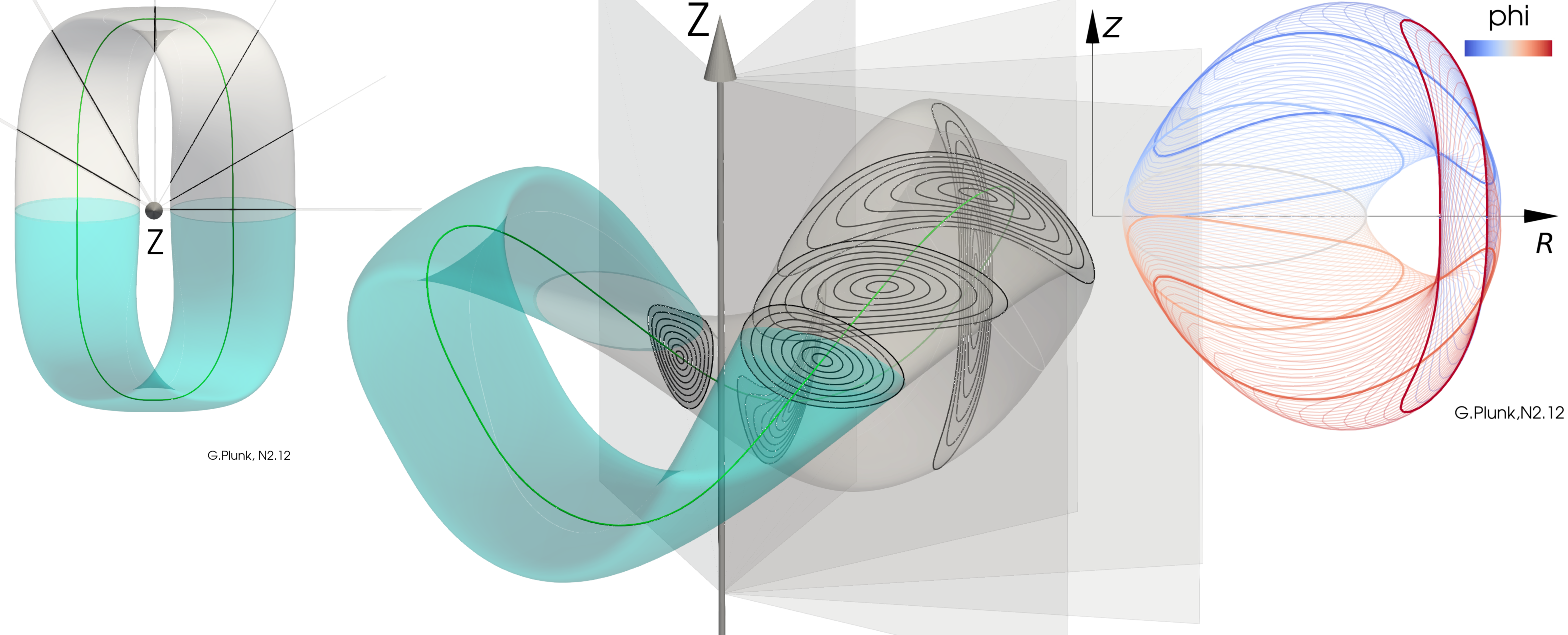}\\[1ex](a) $(R,Z,\phi)$ frame\\[2ex]
    \includegraphics[trim=0 0 0 0,clip,width=0.75\textwidth]{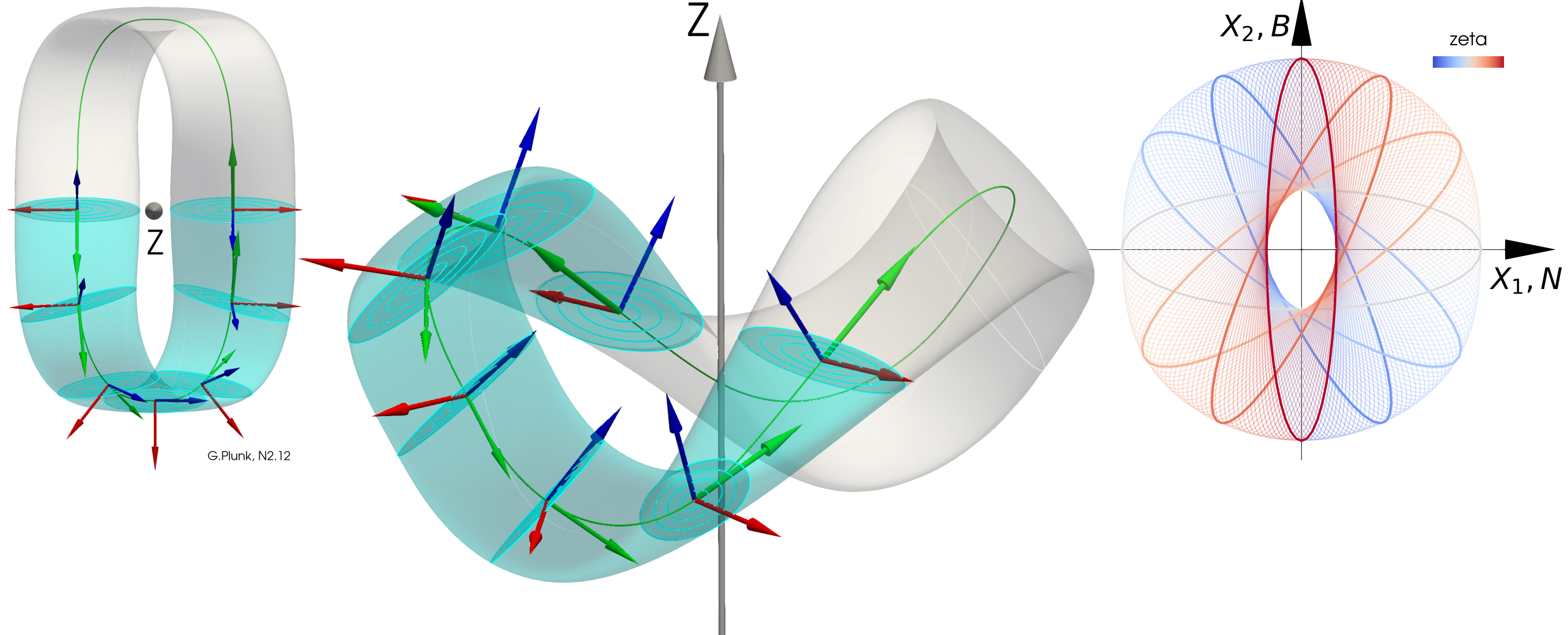}\\[1ex](b) Frenet frame ($T$ green, $N$ red, $B$ blue) \\
    \caption{Comparison of $(R,Z,\phi)$ in and Frenet frame for a two field periodic QI-configuration \texttt{N2-12} \cite{plunk2024-QI}. Top view (left), side view with cross-sections (middle), all cross-sections in respective frame, colored along the toroidal parametrization (right). }
    \label{fig:comparison}
\end{figure}

Using the \GFF{} in an MHD equilibrium solver will enable progress in optimization of new stellarator configurations. In this work we demonstrate, for the case of a highly shaped QI-optimized stellarator, that far fewer degrees of freedom are needed to find a high-quality equilibrium solution, as compared to the solution computed in cylindrical coordinates.  This configuration is depicted in Fig.~\ref{fig:comparison}, showing highly shaped cross-sections in the \RZplanes{}, as opposed to simple ellipses in the \NBplanes{ } of the Frenet frame. 

\begin{figure}[htbp!]
    \centering
    \includegraphics[trim=0 0 0 0,clip,height=0.2\textheight]{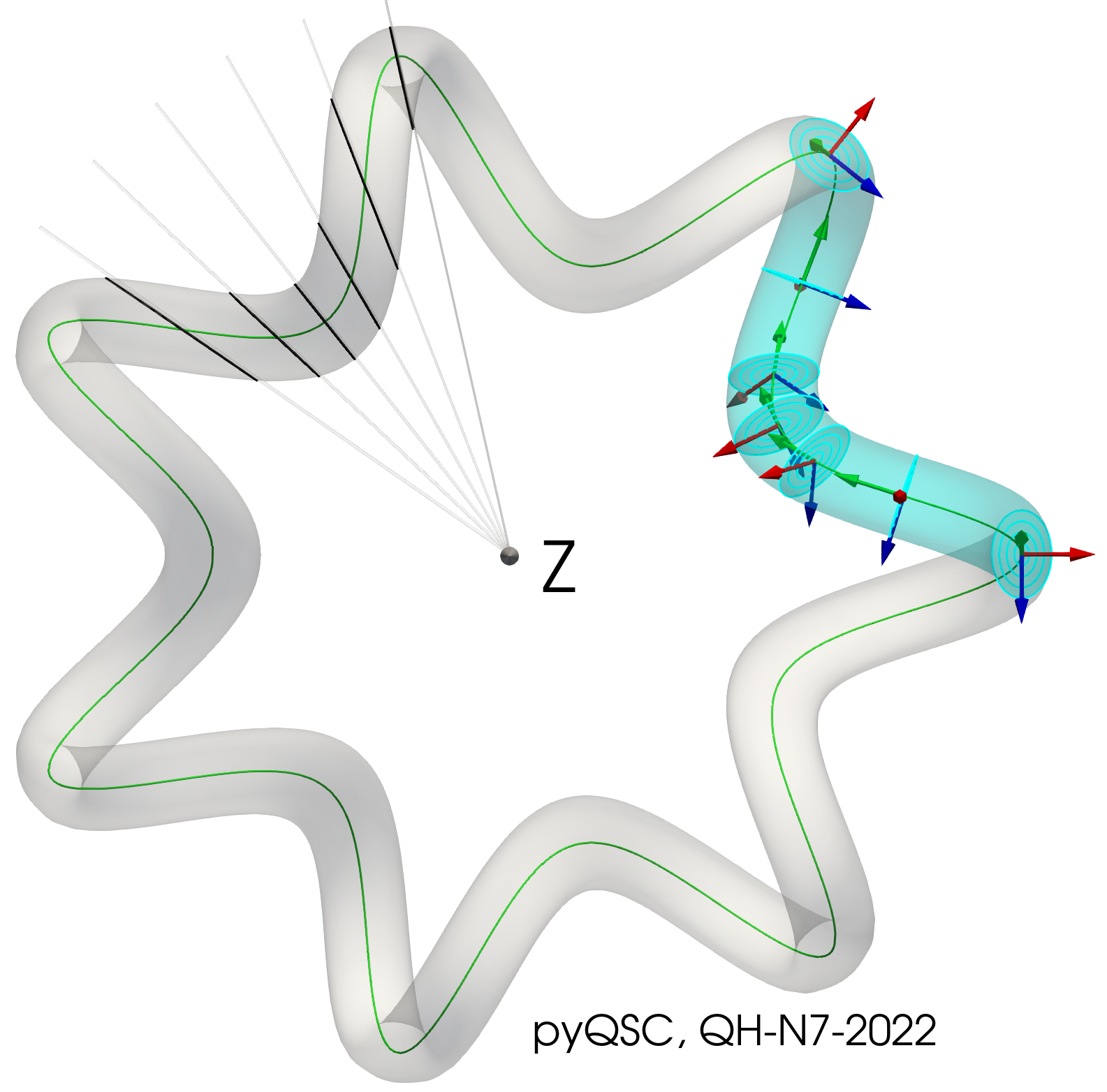}
    \includegraphics[trim=0 0 0 0,clip,height=0.2\textheight]{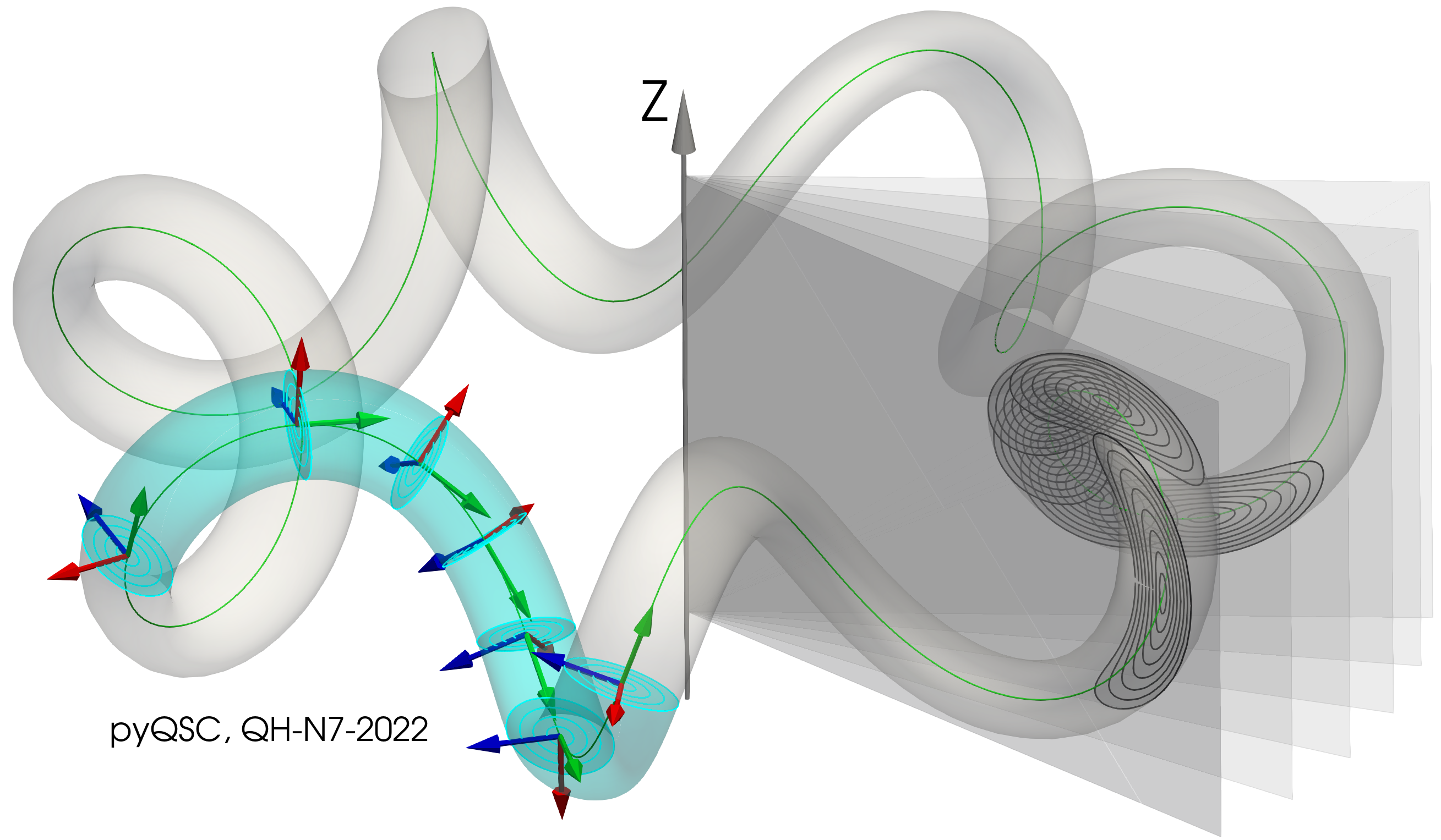}\\[1ex]
    (a) Configuration \texttt{\small 2022 QH nfp7}  generated with pyQSC \cite{landreman-sengupta,landreman-sengupta-plunk}, with 7 field periods \\[2ex]
    \includegraphics[trim=0 0 0 0,clip,height=0.2\textheight]{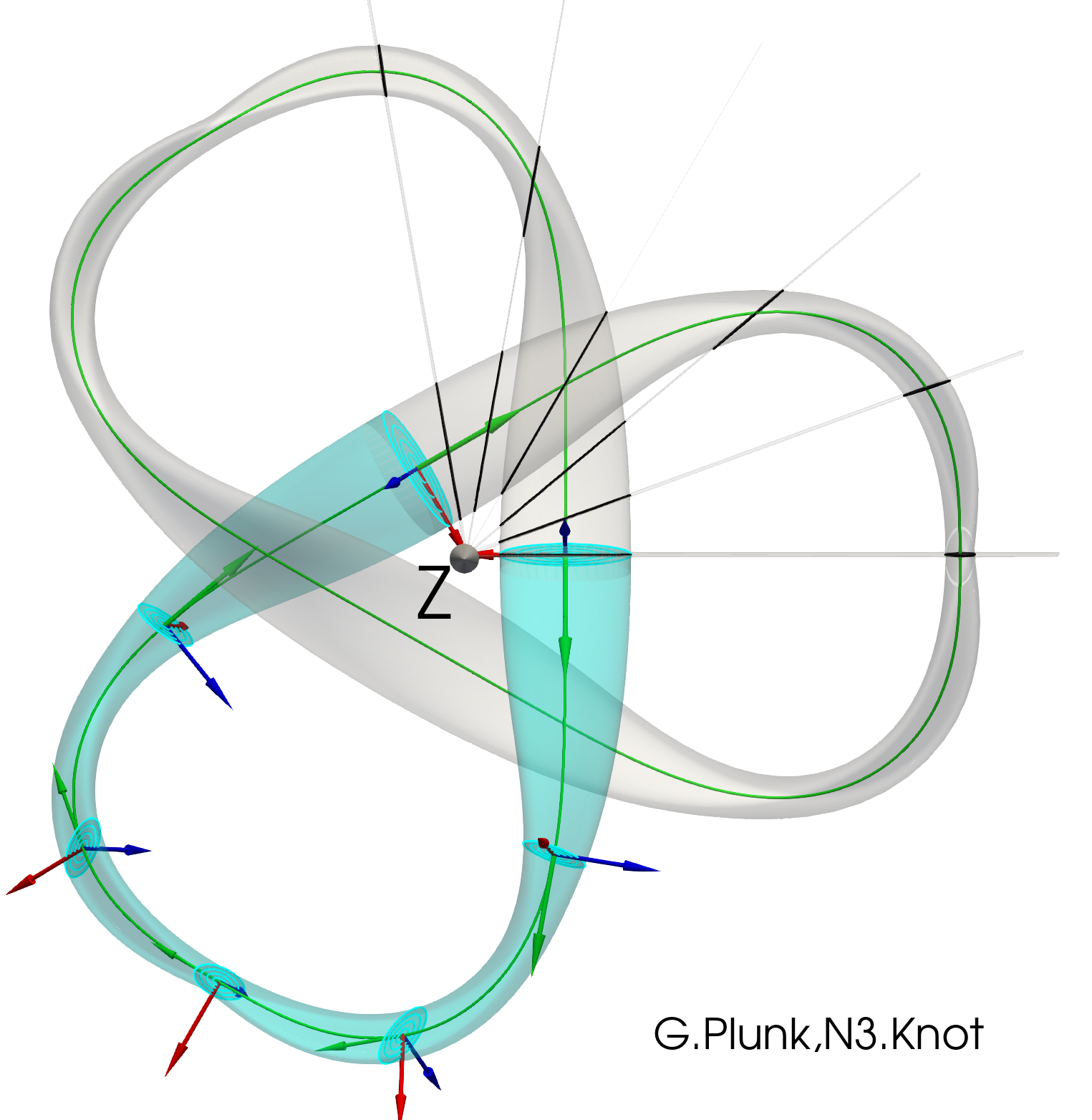}
    \includegraphics[trim=0 0 0 0,clip,height=0.2\textheight]{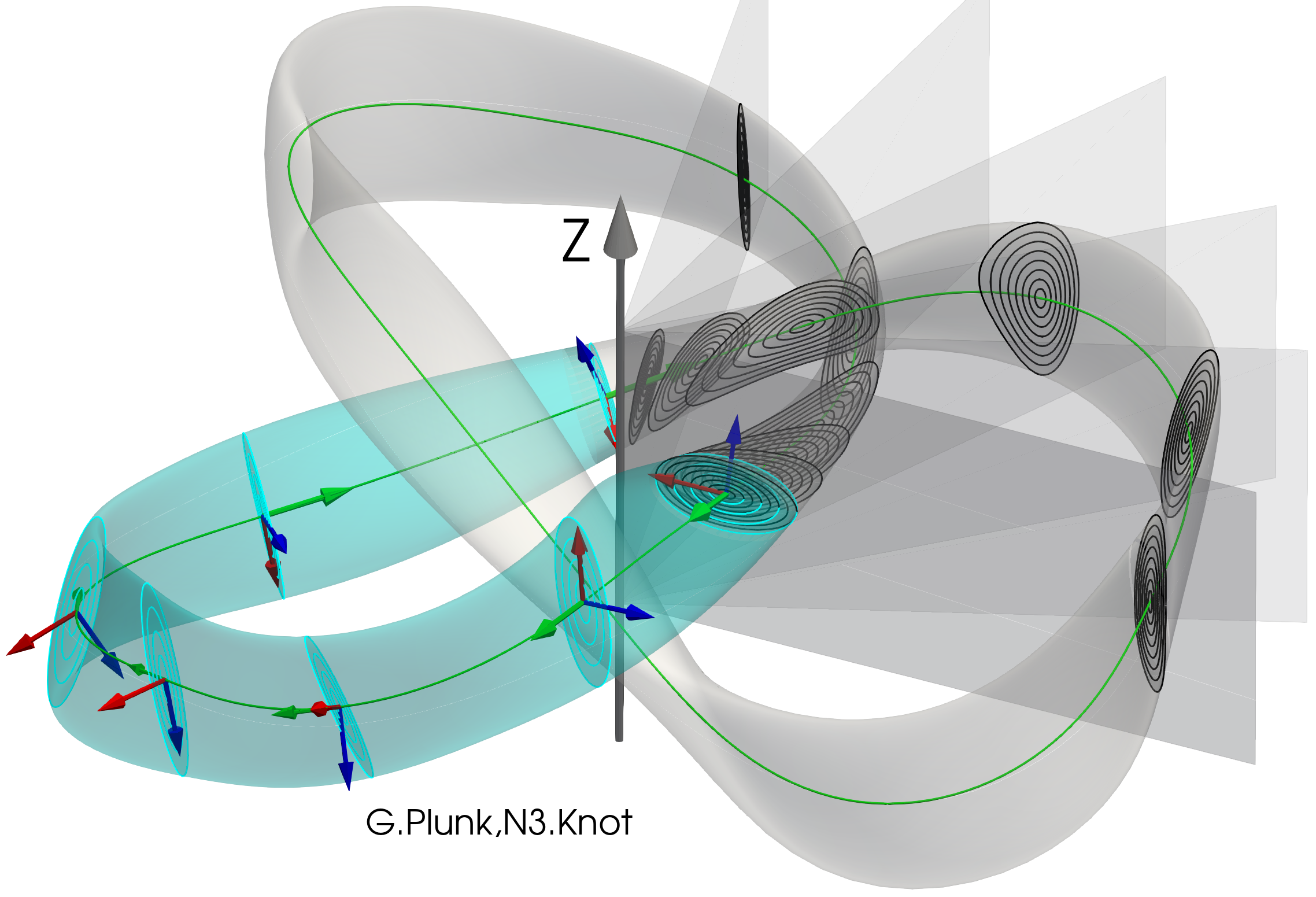} \\[1ex]
    (b) Knotted QI-configuration from NAE \cite{plunk2024-QI}, with 3 field periods.
    \caption{Top view (left) and side view (right) of two configurations. The cross-sections in \RZplanes{ } and in the Frenet frame are shown, one field period is colored green.}
    \label{fig:N7-QH-N3-knot}
\end{figure}

In Fig.~\ref{fig:N7-QH-N3-knot}, examples of a helical magnetic axis and a knotted magnetic axis are shown, where the cross-sections in the Frenet frame remain ellipses. The representation of such configurations in $(R,Z)$ cross-sections is challenging: for the helical case the magnetic axis points in the vertical ($Z$) direction in some places, generating strongly elongated bean shapes, whereas for the knotted case one field period spans $\frac{2}{3}(2\pi)$ in the toroidal angle, and thus would require a range of $\phi \in[0,4\pi)$ for the full configuration. 

Generally, when the 'toroidal' parametrization (we will refer to $\zeta$) is fixed as the geometric angle  $\phi$, it can lead to a non-optimal representation of the surface. 
In contrast, using a Frenet frame, the 'toroidal' parametrization is simply defined by the curve parametrization, and thus can be adapted to the surface shape that one wants to represent.

The outline of the paper is as follows:  \hlchangedrev{In section~\ref{sec:equi}, we start by presenting the necessary derivations for using arbitrary curvilinear coordinates in a 3D MHD equilibrium solver, and introduce the coordinate map that is defined by a \GFF{}. 
Then, we discuss the properties of the conventional Frenet frame and the \GFF{} in section~\ref{sec:gff}, and present two possible ways of constructing it in section~\ref{sec:construction}. 
Finally, we show the numerical results 
for a highly shaped stellarator in section~\ref{sec:results}.}

%%%%%%%%%%%%%%%%%%%%%%%%%%%%%%%%%%%%%%%%%%%%%%%%%%%%%%%%%%%%%%%%%%
\section{\hlchangedrev{MHD equilibrium solver} \label{sec:equi}}
\subsection{MHD equilibrium \hlchangedrev{with flexible curvilinear coordinates}}
In GVEC, we adopt essentially the same approach as in VMEC \cite{hirshman_VMEC_1983} to compute the MHD equilibrium. First, under the assumption of closed nested flux surfaces, the geometry is defined by an embedding map 
\begin{equation}
    f: (\rho,\thet,\zeta) \mapsto (x,y,z)\,,
\end{equation}
mapping from flux-aligned coordinates  (a radial coordinate proportional to the square root of the toroidal magnetic flux, $\rho\sim\sqrt{\Phi}$  and two periodic coordinates $(\thet,\zeta$) to Cartesian coordinates $(x,y,z)$. We decompose the full map $f\coloneqq h\circ \tilde{X}$ with
\begin{align}
    \tilde{X} &: (\rho,\thet,\zeta)\mapsto (q^1,q^2,\hlchangedrev{q^3}) = \big(X^1(\rho,\thet,\zeta),X^2(\rho,\thet,\zeta),\zeta\big) \\
	\hmap &: (q^1,q^2,\hlchangedrev{q^3}) \mapsto (x,y,z)=\hmap(q^1,q^2,\hlchangedrev{q^3})
\end{align}
leaving $X^1,X^2$ as functions of $(\rho,\thet,\zeta)$ to describe the geometry of each cross-section along $\zeta$. The map $\hmap$ is fixed throughout the equilibrium calculation but can be arbitrarily specified by the user at the beginning. Simple examples of $\hmap$ are given by the periodic cylinder $(x,y,z) \coloneqq (q^1,q^2,\zeta)$ and the conventional \hlchanged{cylindrical} representation $(x,y,z) \coloneqq (q^1 \cos\zeta,q^1 \sin\zeta,q^2) = (R\cos\phi, R\sin\phi, Z)$. 

\hladdedrev{It is always assumed that $(x,y,z) = h(q^1,q^2,q^3)$ is an orientation-preserving coordinate transformation, i.e., the Jacobian determinant is strictly positive, $\det (Dh) > 0$. (Here and in the following $D$ denotes the derivative operator.) For the composition $f = h \circ \tilde{X}$ to be defined, the function $\tilde{X}$ must take values in the domain of definition of $h$. This is the only constraint that links $h$ and \hlchanged{$\tilde{X}$}. In addition, we require (independently of $h$) that 
\begin{equation*}
    \det D\tilde{X} = \left(\partial_\rho X^1 \partial_\thet X^2 - \partial_\rho X^2 \partial_\thet X^1\right) > 0, \quad \text{for $\rho > 0$},
\end{equation*}
which is the Jacobian of the transformation $(\rho,\thet) \mapsto (X^1,X^2)$. 
Hence, for any fixed angle $\zeta$, the map $(q^1, q^2) = \big(X^1(\rho,\thet,\zeta), X^2(\rho,\thet,\zeta)\big)$ is an orientation-preserving diffeomorphism from the unit disk, parametrized by polar coordinates $(\rho,\thet)$, into a domain in the $(q^1,q^2)$-plane. By the chain rule, $Df = Dh(\tilde{X}) D\tilde{X}$, we have that $f$ is an orientation-preserving diffeomorphisms away from the polar singularity at $\rho=0$. 
}%hladdedrev
%REMARK: This is not very precise, mathematically. I have a better way to write it, but it requires a complete redefinition of the maps and makes use of some ideas that might not be common knowledge for a physics audience. Is this acceptable to you? %%%YESSS! THANKS OMAR!

The magnetic field is defined by the geometry, an additional angle transformation $\lambda(\rho,\thet,\zeta)$ and the rotational transform profile  $\iota(\rho)=\chi'(\rho)/\Phi'(\rho)$,  with the poloidal magnetic flux $\chi$,  
\begin{equation}
    \mathrm{B}=\frac{1}{\Jac}(b^\thet e_\thet + b^\zeta e_\zeta)\,, \quad b^\thet\coloneqq(\chi'-\Phi' \partial_\zeta\lambda) \,,\quad
     b^\zeta \coloneqq\Phi'(1+\partial_\thet\lambda)\,. 
\end{equation}
Finally, the equilibrium, defined by the flux surface geometry ($X^1,X^2$) and $\lambda$, is found via minimization of total MHD energy:
\begin{equation}
    \begin{aligned}
        W(X^1,X^2,\lambda) &=\int_\Omega \frac{1}{2\mu_0}\mathrm{B}^2 + \frac{p(\rho)}{\gamma-1} dV \\
        &= \int_0^1 \int_0^{2\pi} \int_0^{2\pi} \left(\frac{1}{2\mu_0}\frac{1}{\Jac}\left(b^\thet g_{\thet\zeta}b^\zeta\right)  +\frac{p(\rho)}{\gamma-1} \Jac \right)\,d\rho\, d\thet\, d\zeta \,,
    \end{aligned} \label{eq:energy}
\end{equation}
with a given pressure profile $p(\rho)$ and total toroidal flux $\Phi(1)$. 

As the full map $f$ is decomposed into $f=h\circ \tilde{X}$, the metric terms in \eqref{eq:energy} are computed as
\begin{equation}
    \Jac=\sqrt{G}\left(\ddp{X^1}{\rho}\ddp{X^2}{\thet}-\ddp{X^2}{\rho}\ddp{X^1}{\thet}\right)\,,\quad g_{\alpha\beta}=\ddp{q^{i}}{\alpha}G_{ij}\ddp{q^{j}}{\beta}\,, \quad G_{ij}=\ddp{h}{q^i}\cdot\ddp{h}{q^j}\,, \label{eq:metric_h}
\end{equation}
with $(q^1,q^2,q^3)\coloneqq(X^1,X^2,\zeta)$ and $i,j=1,2,3$ and $\alpha, \beta \in [\rho,\thet,\zeta]$.

\subsection{\hlchangedrev{Curvilinear coordinates constructed from a G-Frame}}
\hladdedrev{We start from an arbitrary parametrized curve $\X(\zeta)$, \hladdedrev{with derivative $d\X(\zeta)/d\zeta =: \Xp(\zeta) \not= 0$}. As we will only represent closed surfaces, the curve must be closed (periodic) and we choose $\zeta\in[0,2\pi]$. We attach two vectors $\Nnew(\zeta),\Bnew(\zeta)$ to the curve, such that they form a right-handed frame with the tangent $\bm T:= \Xp$ as the third vector. We refer to the system of $\bm T,\Nnew,\Bnew$ as \GFF{}.}
\hlchangedrev{
In this paper, we define the coordinate map $h$ in terms of the \GFF{} as}
%As introduced in section~\ref{sec:gff_def},} the map $\hmap$ using the \GFF{} is
\begin{equation}
    \hmap: (q^1,q^2,\zeta) \mapsto (x,y,z)=\X(\zeta)+q^1\Nnew(\zeta)+q^2\Bnew(\zeta) \,. \label{eq:gff_h}
\end{equation}
In order to compute the metric terms in \eqref{eq:metric_h}, we need first derivatives of $\X,\Nnew,\Bnew$  
\begin{equation}
     \qquad \ddp{\hmap}{q^1}=\Nnew\,,\quad \ddp{\hmap}{q^2}=\Bnew\,,\quad \ddp{\hmap}{q^3}=\ddp{\hmap}{\zeta}=\X'+q^1 \Nnew' +q^2 \Bnew' =:\ttilde\,.
\end{equation}
Then the metric tensor and determinant of $\hmap$ amount to
\begin{equation}
    G =  \left[ {\begin{array}{ccc}
       |\Nnew|^2& \Nnew\cdot \Bnew & \Nnew\cdot \ttilde  \\
          &  |\Bnew|^2 & \Bnew\cdot \ttilde  \\ 
        \text{\tiny symm.} &     & |\ttilde|^2 \\
     \end{array} } \right]  \,,\quad \Jh = \ttilde \cdot \left(\Nnew \times \Bnew\right)  \,.\label{eq:metrics_h}
    \end{equation}
When computing variations of the MHD energy with respect to $X^1,X^2$, we also need to compute the derivatives of $\Jh,G$ with respect to $q^1,q^2$, \hlchangedrev{which can be easily obtained.}

\hlchangedrev{Note that, similar to the singularity of the \RZF{} at $R=0$, this coordinate frame becomes invalid where two consecutive \NBplanes{ } intersect. Since the Jacobian is linear in $q^1,q^2$, this singularity appears as a line, at a distance from the curve at a curvature radius $1/\kappa(\zeta)$. } 

\hladdedrev{As this limit depends directly on the given coordinate frame, we can not make a general statement of this restriction. Up to now, we have not yet found a conflicting boundary surface for QI-optimized stellarators, since they tend to align and stretch along the \binormal{ } direction in regions of high curvature of the axis, thus remaining far away from the singular line. A detailed discussion in found in \ref{app:jacobian}.}

\section{The \hlchangedrev{general coordinate} frame \label{sec:gff}}
\subsection{\hladdedrev{The conventional Frenet frame \label{sec:conventional_frenet}}}
In order to motivate why we have introduced a \GFF{}, we first discuss the properties of the conventional Frenet frame.
%In order to discuss the construction of a \GFF{} from a NAE solution,
%In order to motivate the introduction of a \hlchangedrev{general coordinate} frame (\GFF{}), 
%we first introduce the conventional Frenet frame (or Frenet-Serret apparatus, or TNB-frame). 
It is fully defined by an arbitrary parametrized curve $\X(\zeta)$, \hladdedrev{with derivative $d\X(\zeta)/d\zeta = \Xp(\zeta) \not= 0$}. As we will only represent closed surfaces, the axis must be closed (periodic) and we choose $\zeta\in[0,2\pi]$.  
\begin{figure}[htbp!]
    \centering
    \includegraphics[trim=30 0 0 0,clip,width=0.7\textwidth]{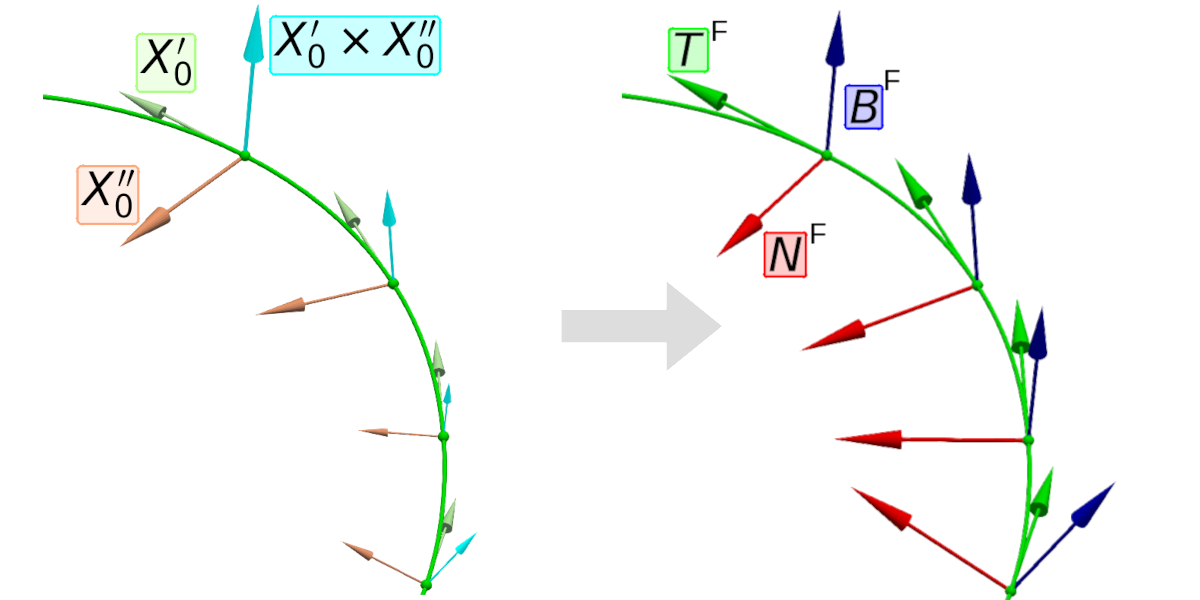} \\
    \caption{Construction of the local Frenet frame from first and second derivatives of the curve $\X(\zeta)$. \hlchangedrev{The normal $\Nfrenet$ scales with the second derivative, so it must be $\neq 0$ to yield a valid Frenet frame.}}
    \label{fig:construct_frenet}
\end{figure}

%We will denote the \emph{\hlchangedrev{two basis vectors that span the cross-sections}  of the \hlchangedrev{general coordinate} frame} by $\bm N$, and $\bm B$. 
\hlchangedrev{To avoid confusion with the previously introduced notation,} we will use the superscript '$F$' when referring to the conventional Frenet frame. Note that $\bm B$ \hlchangedrev{does not refer to the magnetic field.}
The Frenet frame satisfies the following ODE along the curve $\X(\el)$, 
\begin{equation}
			 \dd{}{\el}\begin{pmatrix}
				\Tfrenet\\\Nfrenet\\\Bfrenet
			 \end{pmatrix} = \begin{pmatrix}
				\kappa(\el) \Nfrenet\\-\kappa(\el) \Tfrenet + \tau(\el) \Bfrenet\\- \tau(\el) \Nfrenet
			 \end{pmatrix} \,, \label{eq:frenet_ode}
\end{equation}
with arc-length $\el$, curvature $\kappa$, torsion $\tau$. 
The arc-length is a function of the chosen parametrization of the curve
\begin{equation}
			\el(\zeta) = \int_0^\zeta |\Xp(\tilde{\zeta})| \mathrm{d}\tilde{\zeta} ,\qquad \elp\,(\zeta)  = \dd{\el}{\zeta} =|\Xp(\zeta)|. %\,,\quad \elpp=\frac{\Xpp\cdot\Xp}{\elp}\,. 
            \label{eq:frenet_arclength}
 \end{equation}
 As shown in Fig.~\ref{fig:construct_frenet} and \eqref{eq:frenet_NB}, the  Frenet frame is locally computed from the first and second derivatives of the curve
 \begin{equation}
\Tfrenet=  \dd{\X}{\el} = \dd{\zeta}{\el} \dd{\X(\zeta)}{\zeta}=\frac{\Xp}{|\Xp|}\,,\quad  \Bfrenet=\frac{\Xp\times\Xpp}{|\Xp\times\Xpp|}\,,\quad \Nfrenet=\Bfrenet\times \Tfrenet\,, \label{eq:frenet_NB}
\end{equation}
and the curvature and torsion are
\begin{equation}
    \kappa(\el) = \frac{\left|\Xp\times\Xpp\right|}{|\Xp|^3}\,,\quad \tau(\el) = \frac{(\Xp\times\Xpp)\cdot\Xppp}{|\Xp\times\Xpp|^2}\,.
\end{equation}
\hladdedrev{Equations \eqref{eq:frenet_NB} define a smooth orthonormal frame on the considered curve, provided that $\Xp(\zeta) \times \Xpp(\zeta) \not = 0$. If this quantity vanishes, the \binormal{ } $\Bfrenet$ is indefinite and this can happen if and only if $\kappa(\zeta) = 0$. (We recall that $\Xp(\zeta) \not=0$ per assumption, so that the unit tangent $\Tfrenet$ is always well-defined.) Let us suppose that the curvature vanishes at isolated points only, which appears to be the case in practice. If $\zeta = \zeta_0$ is an isolated zero of the vector-valued function $\Xp(\zeta) \times \Xpp(\zeta)$, we must have $\Xp(\zeta) \times \Xpp(\zeta) = (\zeta - \zeta_0)^n \boldsymbol{u}_0 + O \big(|\zeta-\zeta_0|^{n+1}\big)$, where $\boldsymbol{u}_0 \not=0$ is a constant vector and the integer $n$ is the order of the zero. Then, near $\zeta_0$, we have
\begin{equation*}
    \Bfrenet(\zeta) = \frac{\Xp(\zeta) \times \Xpp(\zeta)}{|\Xp(\zeta) \times \Xpp(\zeta)|} =  \bigg(\frac{\zeta - \zeta_0}{|\zeta - \zeta_0|} \bigg)^n \frac{\boldsymbol{u}_0}{|\boldsymbol{u}_0|} + O\big(|\zeta-\zeta_0|\big),
\end{equation*}
and thus
\begin{equation*}
    \lim_{\varepsilon \to 0^+} \Bfrenet(\zeta_0 \pm \varepsilon) = (\pm 1)^n \frac{\boldsymbol{u}_0}{|\boldsymbol{u}_0|}.
\end{equation*}
We deduce that the standard definition of the \binormal{ } vector given in equations \eqref{eq:frenet_NB} has a discontinuity at an isolated zero of $\kappa$ if the integer $n$ is odd. It is continuous otherwise.
When the discontinuity occurs, it amounts to a change of sign in the direction of the unit vector $\boldsymbol{u}_0 / |\boldsymbol{u}_0|$. Simple examples of this behaviour are given by the curves $\X(\zeta) = (\zeta, \zeta^3, 0)$ and $\X(\zeta) = (\zeta, \zeta^4,0)$ for $\zeta$ in a neighborhood of $\zeta_0 = 0$. In both cases $\zeta=\zeta_0=0$ is an isolated zero of the curvature, with $n=1$ and $n=2$, respectively. In the following, we  refer to these discontinuities as a "flip" of the \binormal{}.  Since, under the assumptions, the unit tangent is smooth, and the Frenet system is right-handed by definition, a flip of the \binormal{ } implies a flip of the normal vector. 
}
%HI OMAR! NICE SECTION!!!
%%

\begin{figure}[htbp!]
    \centering
    \includegraphics[trim=0 0 0 0,clip,width=0.8\textwidth]{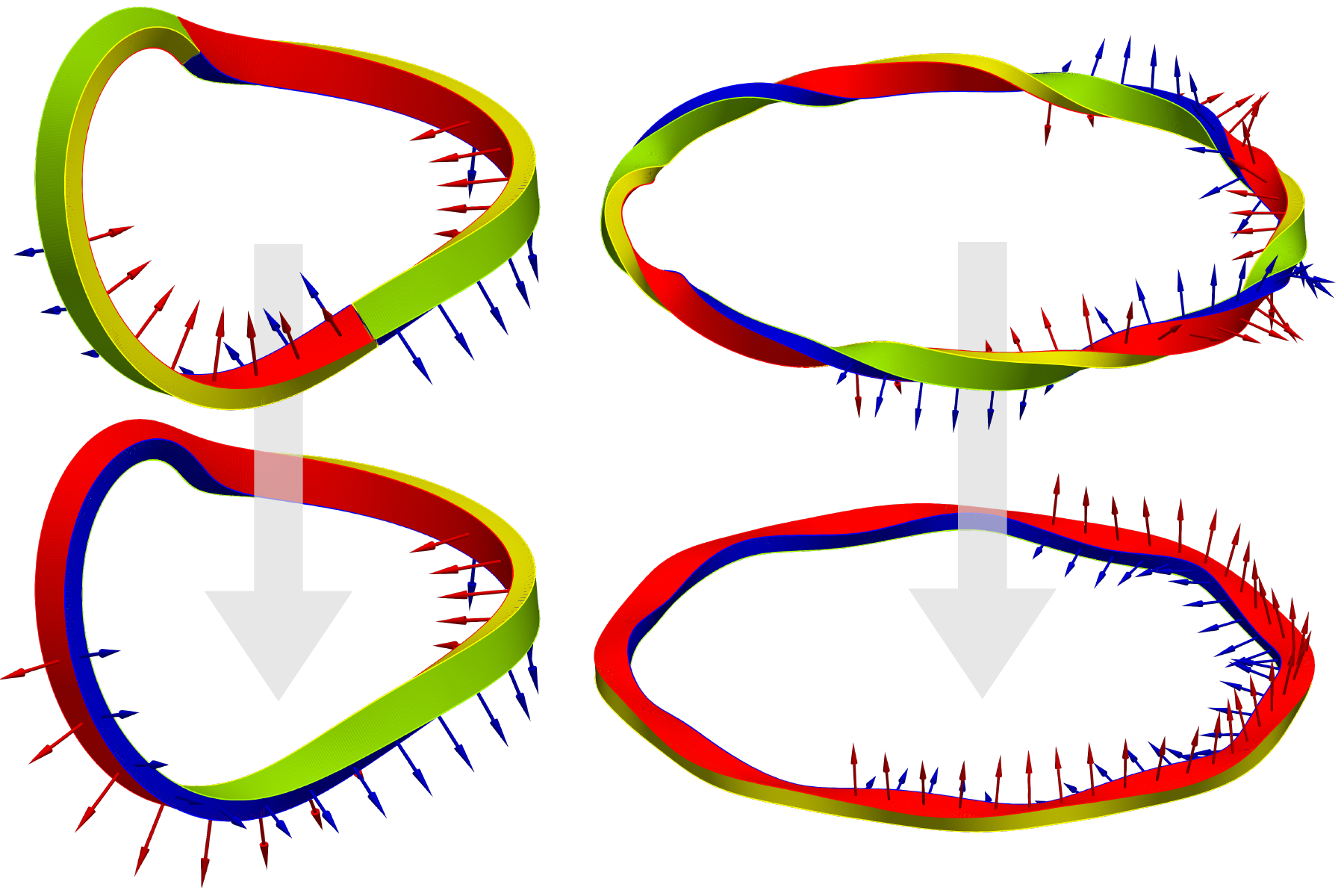}\\
    \caption{Visualization of the $(N,B)$-frame by an axis-centered rectangle swept along the axis, the surface color marks the directions $+N$ in red, $-N$ in green, $+B$ in blue  and $-B$ in yellow. The Frenet frame is undefined at points of zero curvature of the axis, and \hlchangedrev{the normal and \binormal{ } can flip direction. The flip} shown here arises at locations of minimum field strength in quasi-isodynamic stellarators \cite{plunk_landreman_helander_2019} (top left). 
    \hladdedrev{In the \GFF{}, the normal and \binormal{ } are given functions over the curve parameter, allowing to construct} a continuously differentiable frame (bottom left). The Frenet frame  can be twisted, here with a \linknum{ } of $-4$ (top right). The \GFF{} (bottom right) allows to 'untwist', yielding a \linknum{ } of $0$.
    }
    \label{fig:signed_untwist}
\end{figure}

The Frenet frame has the advantage of being defined solely by the curve. However, the following disadvantages exist: 
\begin{enumerate}
    \item The Frenet frame is not defined at points of zero curvature. As stated above, the 
    %TODO "sign of the frame" -> "sign-flip of the normal and bi-normal" ?
    \hlchangedrev{\binormal{ } (and normal) of the frame can flip their direction when crossing such points. This is particularly problematic for stellarator applications since zero curvature points are necessary for quasi-isodynamic
stellarators \cite{plunk_landreman_helander_2019}, and is depicted in the top-left of Fig.~\ref{fig:signed_untwist}.}
    \item Another subtlety is related to the global \linknum{ } of the Frenet frame. The \emph{\linknum{ }} is  the number of rotations of the frame when following it along one turn. First, it is convenient to have zero \linknum{ } for the coordinate frame, for example to ensure that the rotational transform can be determined in a straightforward manner.   \hlchangedrev{Second, it is possible that, after correcting for the frame flips mentioned above, the resulting frame is not periodic.  For example in stellarators with odd field period numbers where there is $1/2$  rotation of the frame in each field period, a sign difference remains after a full toroidal transit.}  An example of a Frenet frame having a \linknum{ } of $-4$ is shown top-right of Fig.~\ref{fig:signed_untwist}.
    %\hladdedrev{Note that there also exist non-integer linking numbers, e.g. $1/2$, which would yield a flip of the normal and \binormal{ } direction at the end of the full turn.}
\end{enumerate}
 
\subsection{\hladdedrev{Defining properties of a G-Frame}\label{sec:gff_def}}
%
%\hladdedrev{In order to mitigate these shortcomings, we propose to provide not only a curve, but also the two basis vectors that span the cross-sections.} 
\hladdedrev{As we have stated above, a \GFF{} is defined by a guiding curve and two basis vectors.} 
\hlchangedrev{Here we formulate the following \emph{requirements}:}
\hlchangedrev{\begin{enumerate}
   \item A smooth closed guiding curve $\X(\zeta)$ is provided, that serves as origin of the cross-sections and its tangent $\Xp(\zeta)$ serves as the first basis vector
   \item The second and third basis vectors $\Nnew(\zeta)$, $\Bnew(\zeta)$ are provided and are \emph{required to be differentiable functions} of the parameter of the curve $\zeta$.
   \item Naturally, the vectors $\Xp$, $\Nnew$ and $\Bnew$ must form a right-handed coordinate system $\Xp \cdot(\Nnew \times \Bnew)>0$, but they \emph{do not need to be orthogonal}.
\end{enumerate}}
Hence, the \GFF{} allows us to effectively undo the twist that comes about in cases of non-zero \linknum{}. A visualization of the resulting \GFF{} is shown in the bottom of Fig.~\ref{fig:signed_untwist}. 

\hladdedrev{Finally, a boundary surface is described by two coordinates $X_b^1(\thet,\zeta),X_b^2(\thet,\zeta)$ and the mapping to 3D Cartesian coordinates, using the basis vectors of the \GFF{}:
\begin{equation} 
\bm x(\thet,\zeta)=\X(\zeta)+X_b^1(\thet,\zeta)\Nnew(\zeta)+X_b^2(\thet,\zeta)\Bnew(\zeta) \,. \label{eq:gff_xyz_surf}
\end{equation}}

\subsection{Interface to the \GFF{} \label{sec:gff_interface}}
In GVEC, we defined an interface to the \GFF{}, that has the following properties:
\begin{enumerate}
    \item It is assumed that $\X(\zeta),\Nnew(\zeta),\Bnew(\zeta)$ are smoothly differentiable along the curve parameter $\zeta$.
    \item The frame is given at a discrete set of points \emph{on the full torus} 
    \begin{equation}
        \zeta_i=\frac{i-1/2}{n_p}(2\pi)\,,\quad i=1,\dots,n_p \label{eq:zeta_i}
    \end{equation}
    \item The number of points $n_p$ must be a multiple of the field periodicity $\nfp$ of the configuration, in order to be able to exactly represent a single field period.
    \item Cartesian coordinates are used to represent the curve position $\X(\zeta_i)$ and normal and \binormal{ } vectors $\Nnew(\zeta_i),\Bnew(\zeta_i)$.
\end{enumerate}

The \GFF{} is not unique and thus there are many ways to construct it. This flexibility in the definition of the \GFF{} is intended to allow users to choose a construction that suits the cases of interest to them. One prominent candidate is the Bishop frame, also called the 'parallel transport frame' ~\cite{bishop_frame,yilmaz_new_bishop} or 'rotation-minimizing frame'~\cite{rotation_minimizing_frame}. 

Another possibility to construct a \GFF{} is to start from a known conventional Frenet frame, identify \hlchangedrev{its flips} and twists, then define a corresponding rotation angle function $\Gamma(\zeta)$ and apply a rigid rotation of the $(\Nfrenet,\Bfrenet)$ plane by
\begin{equation}
    \begin{pmatrix}
       {\bm T}(\zeta)\\\Nnew(\zeta)\\\Bnew(\zeta)
    \end{pmatrix} = \begin{pmatrix}
       1 & 0 & 0\\ 0 &\quad\cos(\Gamma(\zeta)) & \sin(\Gamma(\zeta))\\ 0 & -\sin(\Gamma(\zeta)) & \cos(\Gamma(\zeta))\\
    \end{pmatrix} \begin{pmatrix}
        \Tfrenet(\zeta)\\\Nfrenet(\zeta)\\\Bfrenet(\zeta)
     \end{pmatrix}\,. \label{eq:untwist}
\end{equation}

%%%%%%%%%%%%%%%
\section{\hladdedrev{Constructing the general coordinate frame}\label{sec:construction}}
%%%%%%%%%
\subsection{\hladdedrev{Construction from a NAE solution}\label{sec:gff_cosntruction_frenet}}

\begin{figure}[htbp!]
    \centering
    \includegraphics[width=0.45\textwidth]{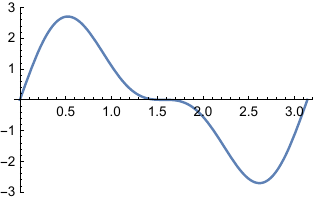}\includegraphics[width=0.45\textwidth]{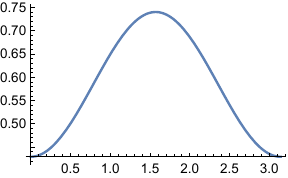}
    \caption{\hladdedrev{Signed curvature $\kappa^s$ and torsion for the magnetic axis of the \texttt{N2-12} configuration.  The horizontal axis is arc-length, which varies from $0$ to $\pi$ over the first field period.}}
    \label{fig:signed-curvature-N2-12}
\end{figure}

\hlchangedrev{The first construction of the \GFF{} starts from} the Frenet frame along the magnetic axis of a QI-optimized NAE configuration~\cite{plunk2024-QI} henceforth called the \texttt{N2-12} configuration. \hlchangedrev{In this case, the Frenet frame} was obtained by solving the ODE~\eqref{eq:frenet_ode} with a given \emph{signed} curvature function, $\kappa^s$, as plotted in Figure \ref{fig:signed-curvature-N2-12}\hlchangedrev{. As a consequence, the \binormal{ } (and normal) directions as defined in equations \eqref{eq:frenet_NB} flip at two locations, namely at $\zeta = \pi/2$ and $\zeta = 3\pi/2$, and the {\em signed} Frenet frame can be defined as $(\TfrenetS, \NfrenetS, \BfrenetS) = (\Tfrenet, s_\kappa\Nfrenet, s_\kappa\Bfrenet)$, where $s_\kappa = \text{sign}(\kappa^s)$. }

\hlchangedrev{Additionally, this signed Frenet frame exhibits a twist of $-\pi$ within a field period. Thus, we introduce a constant rotation of the frame from $0$ to $-\pi$, within each field period, in order to 'untwist' the frame.} Altogether, by using equation \eqref{eq:untwist}, with the rotation function
\begin{equation}
    \Gamma^\text{\texttt{N2-12}}(\zeta) = (s_\kappa-1)\pi/2 - \pi\left(\frac{\nfp}{2\pi}\zeta  - \left\lfloor\frac{\nfp}{2\pi}\zeta\right\rfloor \right) \,,\quad \zeta\in[0,2\pi]\,,
\end{equation}
\hladdedrev{we finally obtain a continuous \GFF{}, without net rotation. The operator $\left\lfloor\frac{\nfp}{2\pi}\zeta\right\rfloor$ yields an integer within each field period.}
 
Furthermore, due to the choice of the point positions \eqref{eq:zeta_i}, the discrete point set excludes the points of zero curvature at $\zeta =[1,3,\dots,2\nfp-1]\frac{\pi}{\nfp} $ as well as the points where the rotation function jumps, at $\zeta =[0,2,\dots,2\nfp]\frac{\pi}{\nfp}$.

\subsection{\hladdedrev{Construction from a boundary surface}\label{sec:quasr}}

\hladdedrev{In this section, we illustrate an alternative construction of the \GFF{}, given a boundary surface only. The construction steps are applied to one example surface from the publicly available QUASR database~\cite{Giuliani_2024_quasr}, with the case \texttt{ID:1942772}\footnote{see \url{https://quasr.flatironinstitute.org/model/1942772}}. It is a QH-optimized stellarator configuration with three field periods, stellarator-symmetry, an aspect ratio of $\sim6$ and mean iota of $1.2$ in vacuum.}

\begin{figure}[htbp!]
    \centering
    \includegraphics[trim=0 0 0 0,clip,width=0.98\textwidth]{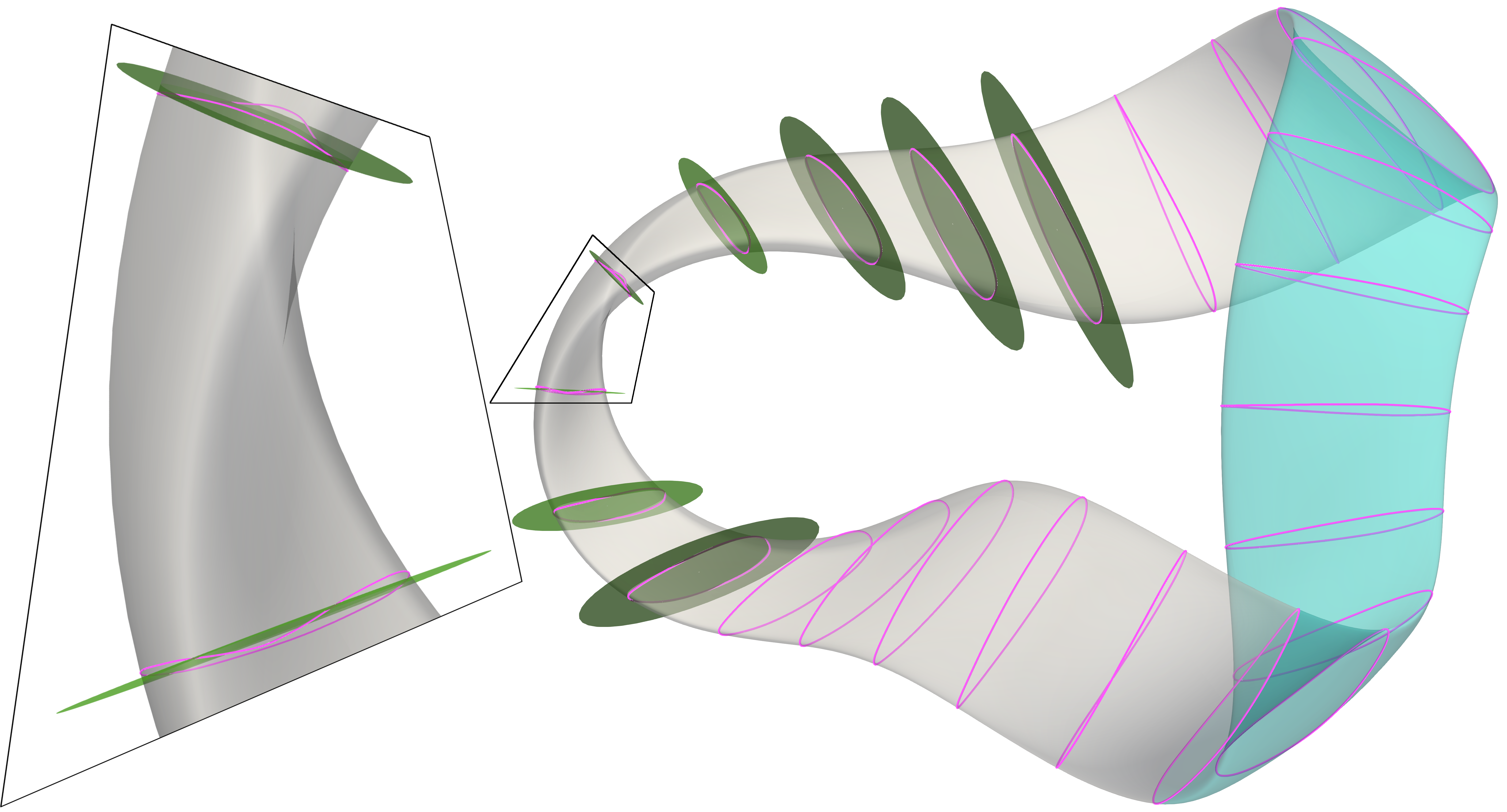}\\
    \caption{\hladdedrev{Visualization of the boundary surface of a three field-period QH stellarator configuration, taken from the QUASR database. Purple are contour lines of constant toroidal parameter $\zeta$. As seen in the zoom-in on the left, the contour lines are non-planar. In green the elliptically-shaped planes are depicted, which can be computed from the first two Fourier coefficients of the $\zeta$ contour lines, from which the  
     \GFF{} can then be deduced.}}
    \label{fig:quasr_init}
\end{figure}

\hladdedrev{The boundary surface is shown in Figure~\ref{fig:quasr_init}. The data format can be read by the \texttt{simsopt} package~\cite{simsopt-joss}, providing a surface geometry object in Cartesian space. 
Note that all \texttt{simsopt} objects from  QUASR are parametrized in \emph{Boozer angles}, which will be taken as the poloidal and toroidal surface parametrization $\thet,\zeta$. The contour lines of the toroidal parameter $\zeta$ are shown as purple curves in Figure~\ref{fig:quasr_init}.}

\hladdedrev{The idea of the proposed construction method is to produce a \GFF{} that follows the orientation of the $\zeta$ contour lines. Hence, if a surface is parametrized in the \RZF{}, the contour lines of $\zeta$ remain in the \RZplanes{}, and the \GFF{} reduces to the \RZF{}. For the \texttt{N2-12} configuration, for which a \GFF{} was constructed in the previous  section~\ref{sec:gff_cosntruction_frenet}, the proposed construction method from its boundary surface would result in the exact same \GFF{}. 
We propose the following steps as a possible construction method from a given boundary surface:}

\hladdedrev{
\textbf{Step 1: Evaluate surface in Cartesian space:}\\
    The target surface, given as $\bm x(\thet,\zeta)$, is sampled on a set of $(\thet_j,\zeta_i)$ positions on the full torus
    \begin{equation}
         \zeta_i=\frac{i-1/2}{n_z}2\pi\,,\quad 
         \thet_j=\frac{j-1/2}{n_t}2\pi\,,\qquad i=1,\dots,n_z ,\quad j=1,\dots,n_t\,.
    \end{equation} 
    If the number $n_z$ is chosen as a multiple of the number of field periods $n_{FP}$, it is possible to reduce the discrete dataset exactly to one field period.}

\hladdedrev{
\textbf{Step 2: Project to surface with elliptical cross-sections:}\\
    The lines of constant $\zeta$ are Fourier transformed in $\thet$, yielding closed curves $\xcont(\thet)$. The zero Fourier mode describes a curve along parameter $\zeta$. It serves as the guiding curve $\X$ of the \GFF{}. 
    \\
    The first Fourier mode of $\xcont(\thet)$ describes a set of planar ellipse curves, centered at the guiding curve $\X$, as shown in green in Figure~\ref{fig:quasr_init}. Note that they are not necessarily orthogonal to the guiding curve tangent vector.}

\hladdedrev{
\textbf{Step 3: Compute the plane of the ellipse cross-sections:}\\
    At each position $\zeta_i$, the second basis vector $\Nnew$ is constructed from $\X(\zeta_i)$ through $\xcont(\thet=0)$, with a unit length. 
    Using the point $\xcont(\thet=\pi/2)$, the unit normal of the ellipse plane $\K$ is computed and yields the third basis vector $\Bnew=(\K\times\Nnew)$ being orthogonal to $\Nnew $ and of unit length.}

\begin{figure}[htbp!]
    \centering
    \includegraphics[trim=0 40 0 20,clip,width=0.9\textwidth]{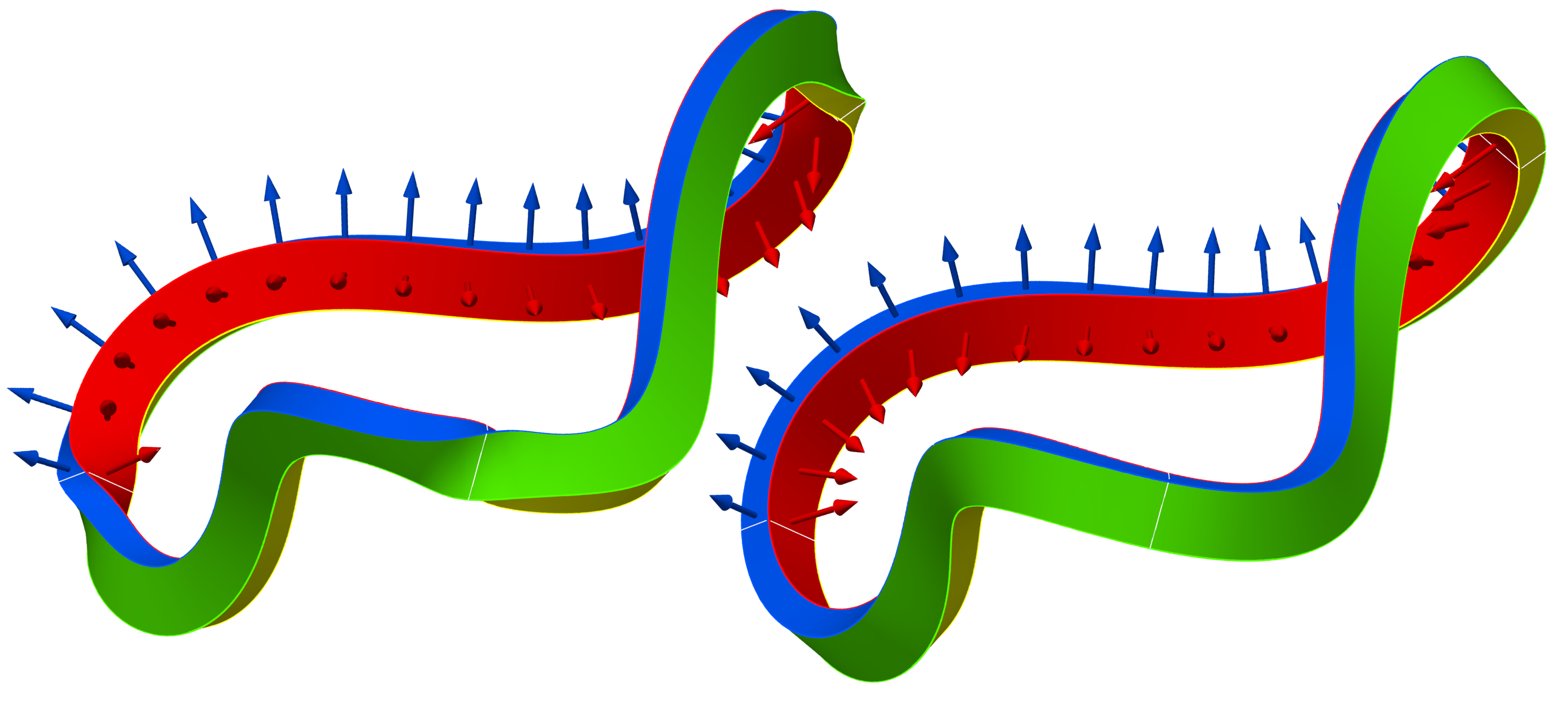}\\
    \hspace{0.3\textwidth}\includegraphics[width=0.55\textwidth]{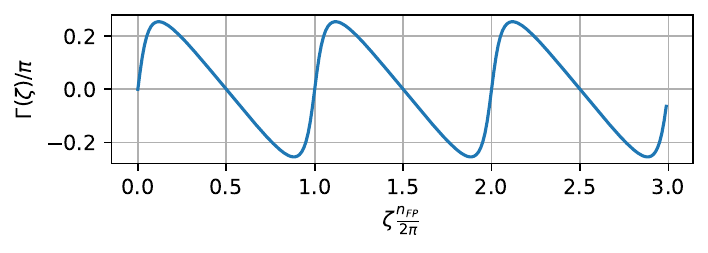}
    \caption{\hladdedrev{Visualization of the \GFF{} computed from the boundary surface. Left:  Initially, a strong twist appears at the end of each field period (white lines). Right: Final smooth frame obtained with the rotation function $\Gamma(\zeta)$ (bottom). The frame is visualized by an axis-centered rectangle swept along the axis, the surface color marks the directions $+N$ in red, $-N$ in green, $+B$ in blue  and $-B$ in yellow.}}
    \label{fig:quasr_untwist}
\end{figure}

\hladdedrev{
\textbf{Step 4: Guarantee a smoothly varying frame:}\\
    In the third step, the direction of the second basis vector $\Nnew$ was fixed to the surface parameter $\thet=0$. The resulting \GFF{} is shown on the left of Figure~\ref{fig:quasr_untwist}. 
    It has a zero \linknum{}, but also twists strongly at the end of each field period. \\
    Knowing that in each cross-section, we have a two-dimensional centered rotated ellipse, we can relate its Fourier coefficients to the  rotation angle of the ellipse, see details in \ref{app:ellipse}. Now we can obtain an improved frame by applying a rigid rotation to the basis vectors $\Nnew$ and $\Bnew$ by \eqref{eq:untwist}.  This rotation, specified by the smooth rotation angle function $\Gamma(\zeta)$, is designed such that the ellipse rotates at a constant rate as measured in the final frame.
    The rotation angle function is shown at the bottom of Figure~\ref{fig:quasr_untwist}. \\
    In the final frame, the rotating ellipse is represented by a single poloidal and a single toroidal Fourier mode, 
    thus smoothly following the target surface, as shown on the right of Figure~\ref{fig:quasr_untwist}.}

\hladdedrev{
\textbf{Step 5: Cut the original surface:}\\
    The final step is to cut the surface with the \NBplanes{ } of the \GFF{}. For each cross-section, we compute the intersection of all curves $\bm x(\thet_j,\zeta)$, and  deduce their coordinates $X^1(\thet_j,\zeta_i),X^2(\thet_j,\zeta_i)$ in the \GFF{}.}

\begin{figure}[htbp!]
    \centering
    \includegraphics[trim=40 10 90 20,clip,width=0.98\textwidth]{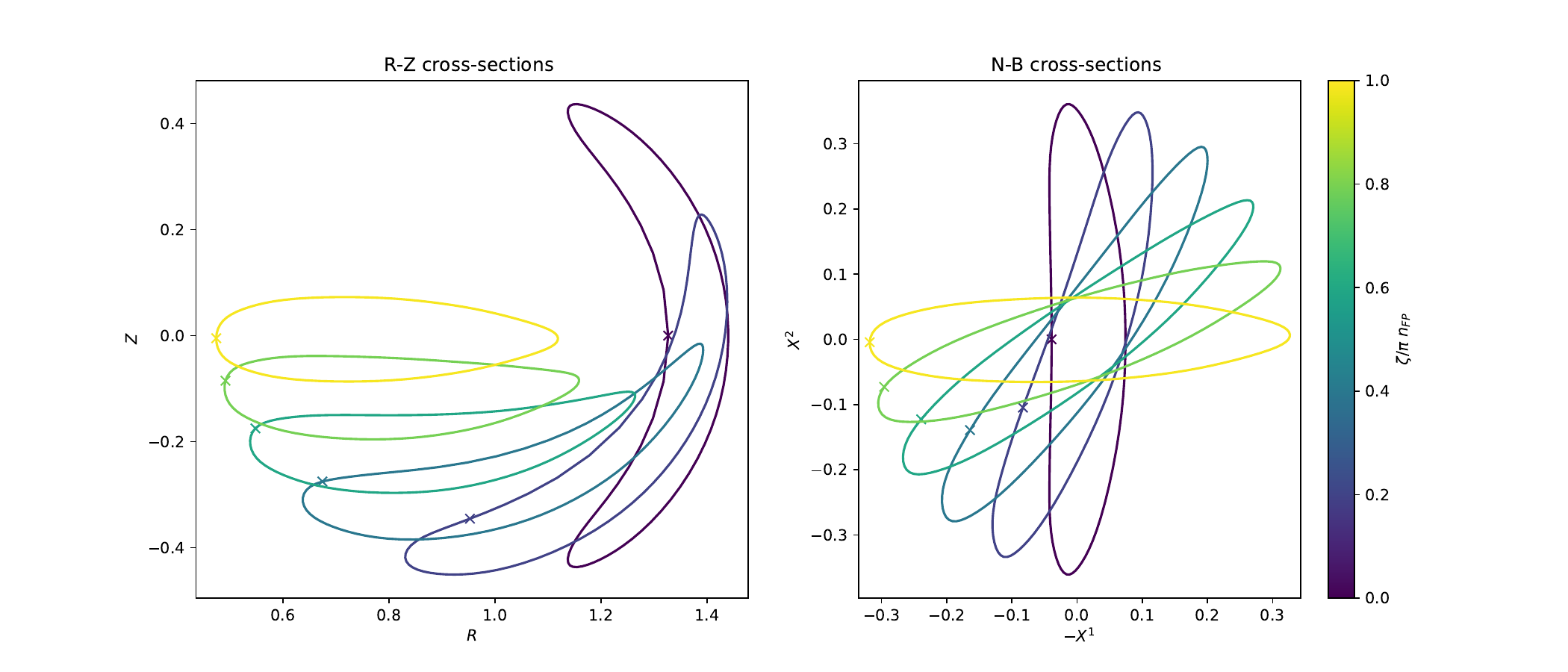}\\
    \caption{Comparison of cross-sections of the boundary surface over half a field period, in the \RZF{} (left) and in the \GFF{}  computed from the boundary surface (right). The cross indicates the point of $\thet=0$.}
    \label{fig:quasr_cross_sections}
\end{figure}

\hladdedrev{The final result is shown in Figure~\ref{fig:quasr_cross_sections}, where we compare the cross-sections of the boundary surface in the \RZF{} to the cross-sections in the \NBplanes{ } of the \GFF{}. We observe that the cross-sections have been greatly simplified. Note that the construction preserves the stellarator symmetry of the target surface. Also note that we use the same algorithm of step 5 to compute the cross sections in the \RZF{}.}

\begin{figure}[htbp!]
    \centering
    \includegraphics[trim=40 10 0 20,clip,width=0.98\textwidth]{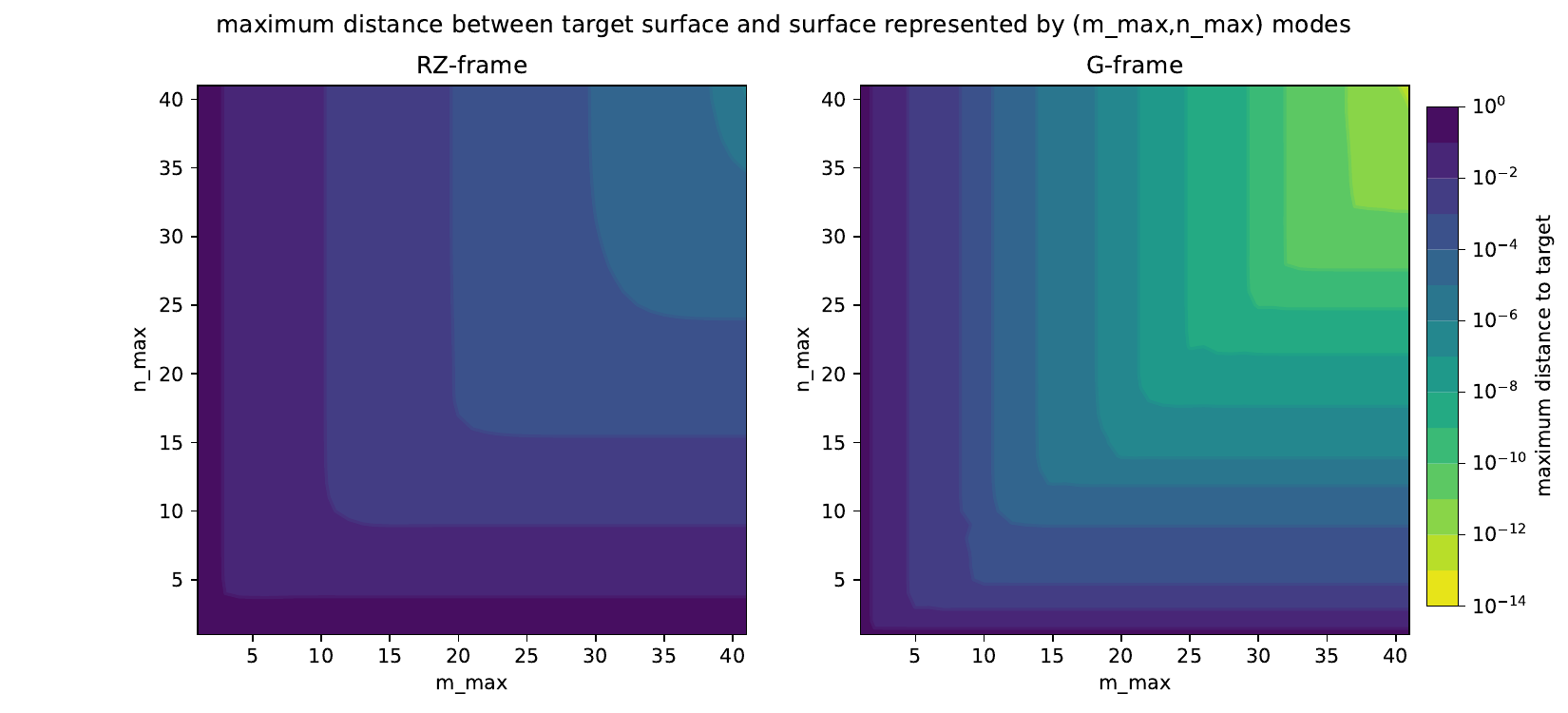}\\
    \caption{\hladdedrev{Comparison of the maximum distance error to the target surface ($|{\bf x}_\text{target}-{\bf x}_{(m,n)_\text{max}}|$ in Cartesian coordinates), after applying a Fourier filter with $(m,n)_\text{max}$ modes. Left:  the coordinate variables of the \RZF{} are represented with $(R,Z)_{(m,n)_\text{max}}$ modes. Right:  the coordinate variables of the \GFF{} are represented with $(X^1,X^2)_{(m,n)_\text{max}}$  modes, showing that fewer modes are needed for the same error. The target surface is represented by  $(m,n)_\text{max}=(40,40)$ modes. }}
    \label{fig:quasr_spectrum_vs_accuracy}
\end{figure}

\hladdedrev{Let us analyze the two resulting surface representation in more detail. In both frames, we represent the target surface by a grid of $81 \times 81$ points per field period, leading to a maximum mode number of $(m,n)_\text{max}=(40,40)$.
We take this as the reference target surface, and apply a Fourier cut-off filter to the corresponding coordinate variables $R,Z$ and $X^1,X^2$, respectively and sample the surface at the same number of points. We can then measure the largest distance between the filtered surface and the target surface, in Cartesian space. We plot this error over the maximum number of poloidal and toroidal modes that were used for the filter. The results are shown in Figure~\ref{fig:quasr_spectrum_vs_accuracy}. It can be clearly seen that the error decays much quicker for the \GFF{}. 
}

\begin{figure}[htbp!]
    \centering
    \includegraphics[trim=40 10 80 20,clip,width=0.98\textwidth]{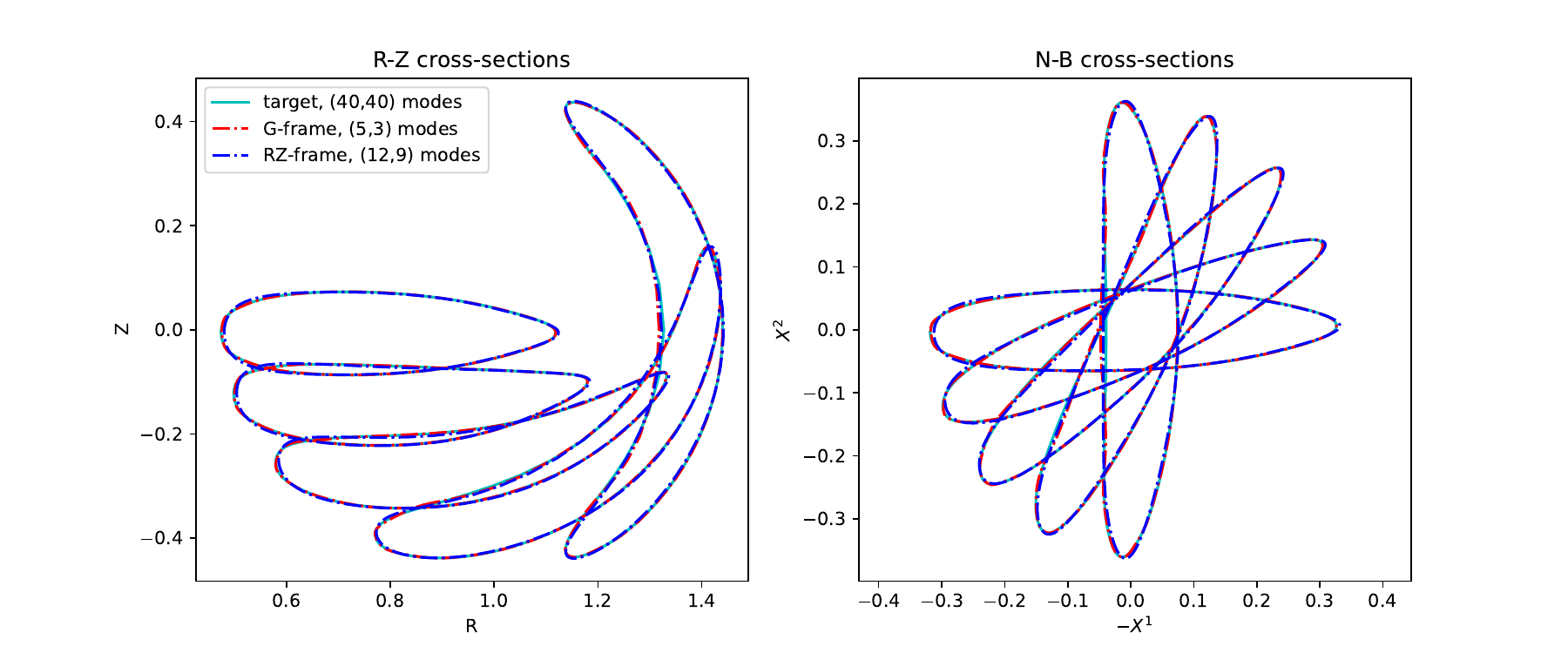}\\
    \caption{\hladdedrev{Comparison of cross-sections of the boundary surface over half a field period. The cross-sections are computed in  \RZplanes{ } (left) and \NBplanes{ }  (right). Shown are cross-sections of target boundary surface (cyan), along with two  Fourier-filtered solutions that have the same maximum distance error of $\sim1\%$. For $R,Z$ as coordinate variables, $(m,n)_\text{max}=(12,9)$ modes are needed (in blue) and using $X^1,X^2$ as coordinate variables in the \GFF{} only $(m,n)_\text{max}=(5,3)$ modes are needed (in red).}}
    \label{fig:quasr_cross_sections_maxdist}
\end{figure}

\hladdedrev{
Finally, we choose a filtered solution in both frames that has the same error, a maximum distance of $\sim 1 \%$. We compute the cross-sections of the target surface and the two filtered solutions. Both types of cross-sections, in  \RZplanes{ }  and \NBplanes{}, are shown in Figure~\ref{fig:quasr_cross_sections_maxdist}. We see that we obtain a similar surface, but in the \GFF{}, the $X^1,X^2$ coordinates only require $(m,n)_\text{max}=(5,3)$ modes ($42$ coefficients in total), whereas in the \RZF{}, the $R,Z$ coordinates require $(12,9)$ modes ($247$ coefficients in total) to reach this accuracy. 
}

\section{Application to a highly shaped stellarator \label{sec:results}}
In this section, we will run GVEC with the new feature, using the \GFF{} for the map $\hmap$. We compare the results with the GVEC solution computed with $\hmap$ given in cylindrical coordinates, which will be referred to as the '\RZF{}' in the following. 
The data of all simulations in this section is openly available at \hlchangedrev{\url{https://doi.org/10.5281/zenodo.14714598}}.

As a challenging example, we start from the first order NAE solution of the two field periodic QI-configuration \texttt{N2-12} \cite{plunk2024-QI}, already shown in Fig.~\ref{fig:comparison}b. 
The average minor radius $\bar{a}$ is computed from the cross-section area in the \NBplane{}, averaged over the arc-length parameter $\ell$ of the axis \eqref{eq:frenet_arclength}, as
\begin{equation}
    \pi\bar{a}^2\coloneqq \frac{1}{\ell(2\pi)}\int_0^{2\pi}
     |\Nnew\times \Bnew|\left |\, \frac{1}{2}\int_0^{2\pi} 
    \left( \ddp{X_b^2}{\thet}X_b^1-\ddp{X_b^1}{\thet}X_b^2 \right) d\thet \,\right | \ell'(\zeta) d\zeta\,,
\end{equation}
using the boundary coordinates $X_b^1(\thet,\zeta),X_b^2(\thet,\zeta)$.
The \GFF{} and the boundary are given by the NAE solution. For the \texttt{N2-12} case, the axis length is $\ell(2\pi) = 2\pi$ meters and the average minor radius is $\bar{a}=0.186756$ meters.
The iota profile is constant and set to the on-axis value $\iota=0.7$ and zero pressure is used, as assumed for the NAE. The average of the magnetic field strength on axis is $B_0=1$ Tesla.

\begin{figure}[htbp!]
    \centering
    \includegraphics[width=0.98\textwidth]{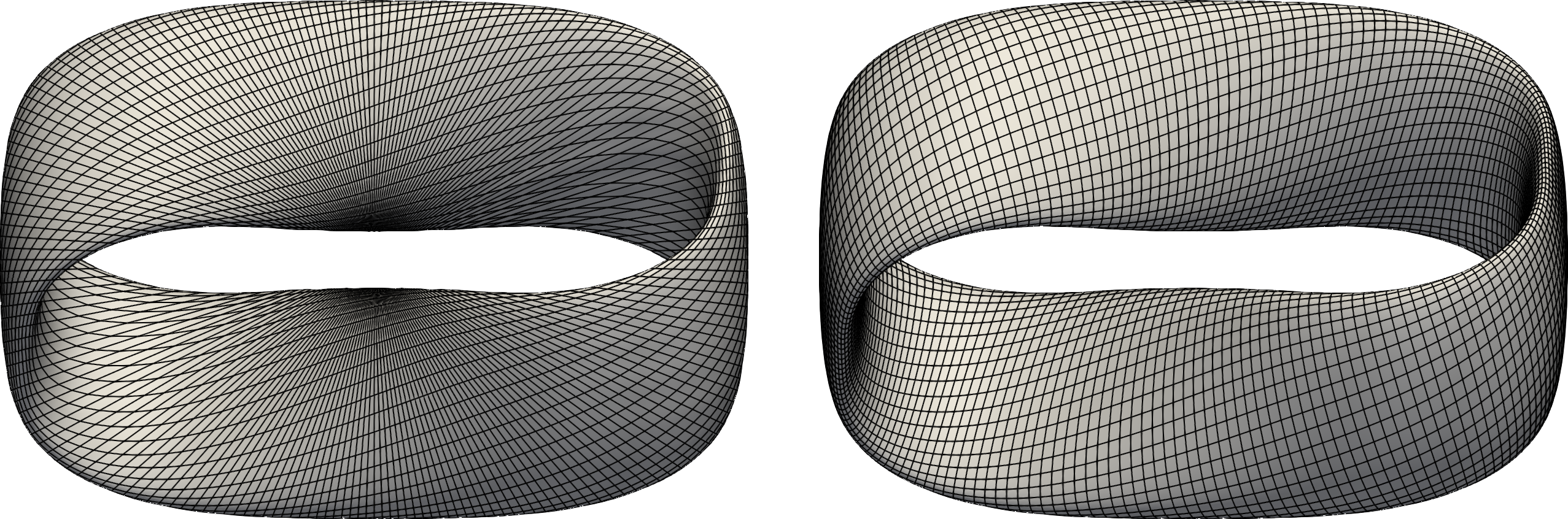}
    \caption{Top view of the boundary parametrization of the 'N2-12' configuration, left using the \RZF{} with  $(m,n)_\text{max}=(10,15)$ and right the \GFF{} with $(m,n)_\text{max}=(2,10)$. \hlchangedrev{The grid is generated by isocontours of the surface parametrization, with $40$ levels poloidally and $200$ levels toroidally.}} 
    \label{fig:boundary_param}  
\end{figure}

In order to run the same case in the \RZF{}, the boundary must be transformed by finding the cross-sections on planes $\phi=$constant of the geometric toroidal angle $\phi$. The change in parametrization is depicted in Fig.~\ref{fig:boundary_param}. The Fourier resolution of the boundary in the \GFF{} is $(m,n)_\text{max}=(2,10)$, whereas in the \RZF{}, a higher resolution of $(m,n)_\text{max}=(10,15)$ is needed to accurately represent the same boundary surface.

\begin{figure}[htbp!]
    \centering
    \includegraphics[width=0.98\textwidth]{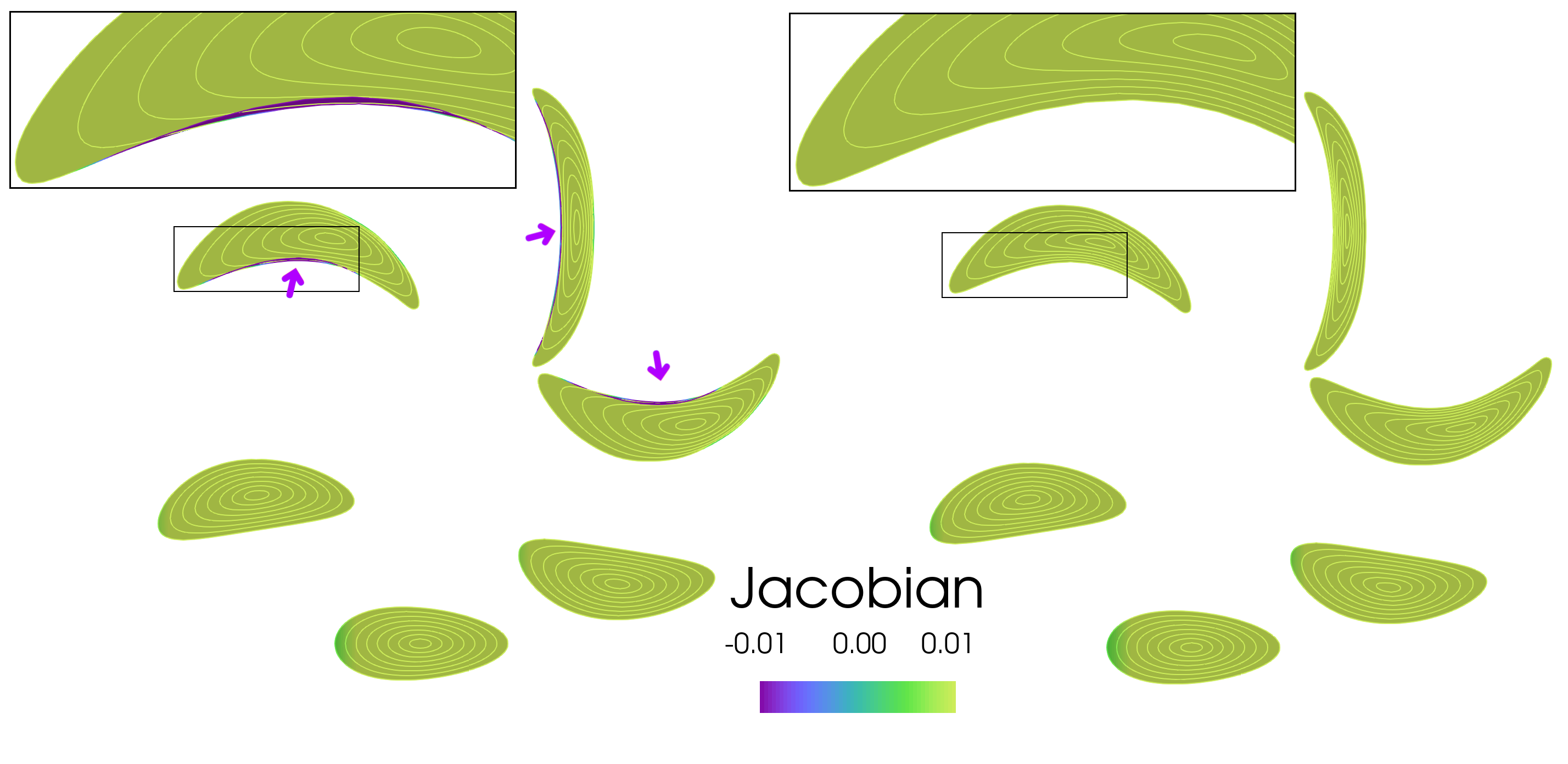}
    \caption{Cross-sections in the \RZF{} over half field period of the 'N2-12' configuration. On the left: Initial guess of the $\rho$ contours, producing an invalid initial map with negative Jacobian close to the boundary (arrows) and a zoomed-in cross-section. On the right: the same view of the valid initial map after application of an action principle~\cite{tecchiolli_constructing_2024}.} 
    \label{fig:action_min}  
\end{figure}

Since cross-sections in the \RZF{} are strongly shaped, the initial guess in GVEC produces a invalid map (negative Jacobian determinant), prohibiting the solver to start the minimization. The same problem occurs in VMEC. Fortunately, we were able to find a valid initial guess, using a recently proposed action principle \cite{tecchiolli_constructing_2024} that we have implemented in GVEC, see Fig.~\ref{fig:action_min}. 

We choose the following resolution parameters for the equilibrium runs in GVEC: 
For the  \GFF{}, we choose a radial resolution of $2$ B-spline elements of degree $5$ and a Fourier resolution of $(m,n)_\text{max}=(2,10)$. 
For the \RZF{}, we choose a radial resolution of $8$ B-spline elements of degree $5$ and a Fourier resolution of $(m,n)_\text{max}=(10,15)$. 
The  minimization in GVEC converged in $800$ iterations with the \GFF{} and in  $16,000$ with the \RZF{}, due to the higher resolution. 

\begin{figure}[htbp!]
    \centering
    \includegraphics[width=0.85\textwidth,trim=20 350 30 70,clip]{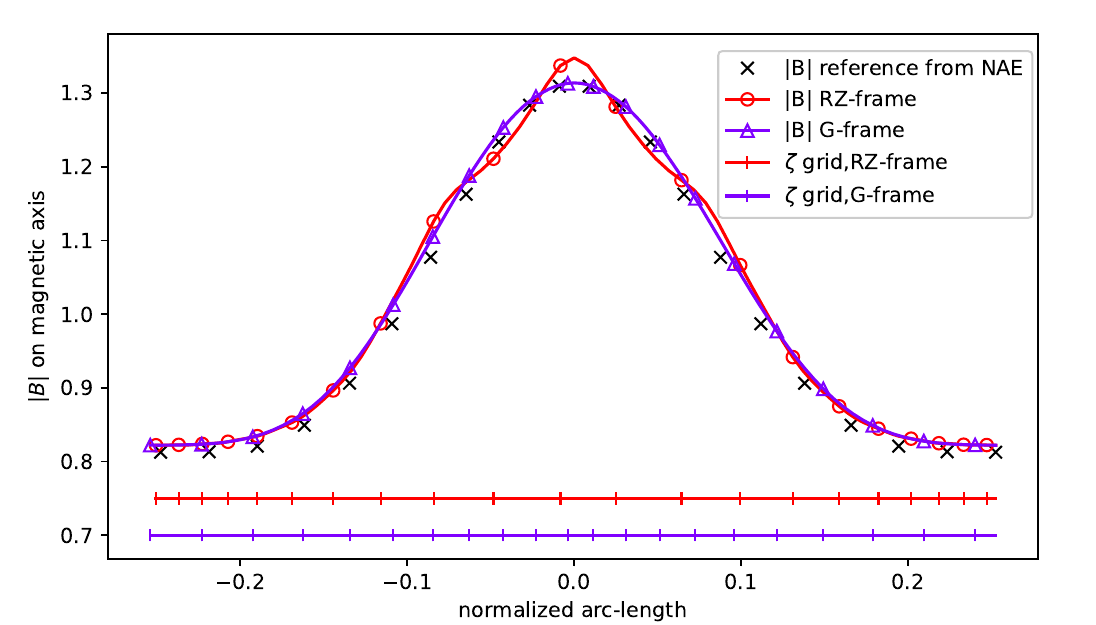}\\
    \caption{Comparison of the magnetic field strength along the magnetic axis of the equilibrium solutions using the \RZF{} and the \GFF{}, and the reference solution of the NAE. \hladdedrev{Note that the normalized arc-length is used along the x-axis, and the parametrization is shown as a '$\zeta$-grid'}} 
    \label{fig:compare_normB}  
\end{figure}

\begin{figure}[htbp!]
    \centering
    \includegraphics[width=0.9\textwidth,trim=0 0 0 0,clip]{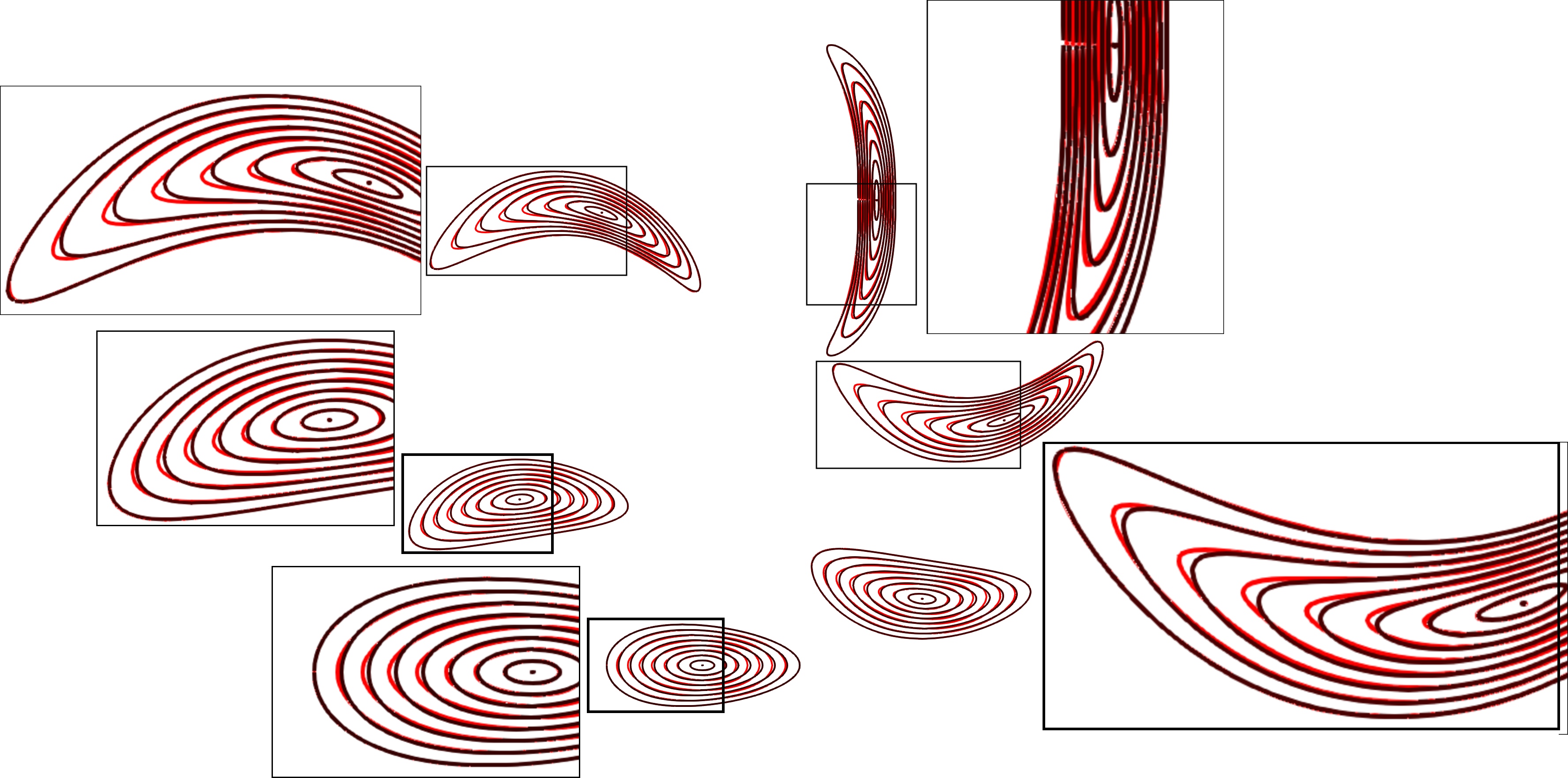}\\
    \caption{Comparison of the equilibrium solution with the \GFF{} in black and the \RZF{} in red, showing cross-sections of flux surfaces in \RZplanes{ } over a half field period. The frames show zoomed in views.} 
    \label{fig:compare_RZ_planes}  
\end{figure}

From the solution of the NAE, we know the magnetic field strength along the axis, which we will use as a reference. In Fig.~\ref{fig:compare_normB}, we compare to the magnetic field strength along the magnetic axis of the equilibrium solutions with the ~\GFF{} and \RZF{}. 
\hlchanged{We choose to plot over the \emph{normalized arc-length} of the magnetic axis. We also plot
the points along a '$\zeta$-grid', showing that the resolution when using the \GFF{} is distributed towards the high-field region, unlike the \RZF{},
having the lower resolution in that region.}
It is clearly seen that the equilibrium solution with the \GFF{} is very close to the reference solution, whereas, even with a higher resolution, the equilibrium solution with the \RZF{} struggles to accurately represent the peak. In Fig.~\ref{fig:compare_RZ_planes}, cross-sections of the flux surfaces show a clear difference of the equilibrium solution, especially in the strongly shaped cross-sections.  

It is expected that the full 3D MHD equilibrium solution using the boundary of the NAE is different from the solution of the NAE. For example, the euclidean distance of the magnetic axis to the axis of the \GFF{} is on the order of $10^{-4}$. But we also know that the 3D MHD equilibrium solution should converge to the solution of the NAE when the boundary is evaluated at a smaller radius. In Fig.~\ref{fig:convergence_NAE}, we demonstrate the  convergence, using GVEC with the \GFF{} and the same resolution as before, but with the initial average minor radius scaled by a factor of $0.1$ and $0.01$. 

\begin{figure}[htbp!]
    \centering
    \includegraphics[width=0.85\textwidth,trim=20 50 30 370,clip]{pics/comparison_Bnorm__r_convergence_over_arclength.pdf}
    \caption{Convergence of the equilibrium solution with the \GFF{} to the solution of the NAE.} 
    \label{fig:convergence_NAE}  
\end{figure}

\section{Conclusion}
In this work, we proposed a new and flexible approach of representing the geometry of 3D MHD equilibria of strongly shaped stellarators. Motivated by the discovery of new configurations found by means of the near-axis expansion (NAE), we adopt the idea of an axis-following frame with cross-sections normal to the axis, instead of the usual cylindrical coordinates ('\RZF{}'). 
We show  the deficiencies of the conventional Frenet frame, such as abrupt \hlchangedrev{flips of the basis vector directions} or unnecessary twists and we introduce a \hlchangedrev{general coordinate} frame (\GFF{}),
which overcomes these deficiencies. We describe the steps needed for implementation of the \GFF{} in the 3D MHD equilibrium solver GVEC.

Using the \GFF{} for the particular case of a two-field-periodic QI-configuration, we show that far fewer degrees of freedom are needed for the representation of the boundary, compared to the \RZF{}. Consequently, we found that an accurate equilibrium solution, obtained in GVEC with the \GFF{}, also needs less resolution and less iterations to converge, compared to the equilibrium solution with the \RZF{}.

\hladdedrev{We want to stress that the approach is not inherently reliant on a NAE solution. Therefore, we demonstrate that it is also possible to construct the \GFF{} from a boundary surface, without knowledge of the magnetic axis or its Frenet frame.}

Computing 3D MHD equilibria in a \RZF{} has strong limitations on the 3D shapes it can represent, artificially restricting the space of possible shapes in stellarator optimization. \hladdedrev{Recently, the first optimization of a Figure-8 shaped stellarator was conducted using GVEC  \cite{plunk2024figure8}, which was made possible by using the \GFF{} for the boundary surface representation, based on a quasi-isodynamic NAE solution.}

\hlchangedrev{We believe that a frame that follows the shape of the surface, such as the \GFF{}, will allow the representation of stellarators of virtually any shape with minimal number of degrees of freedom, and thereby open paths for progress in stellarator optimization. We show with two examples on how to construct the \GFF, and hope to gain more insight on how to optimally represent any boundary surface in the future.}

\ack 
The authors would like to thank our colleagues at the Numerical Methods in Plasma Physics division at IPP Garching and the Stellarator Theory group at IPP Greifswald for the insightful discussions and support in this endeavor. We would also like to thank Matt Landreman for a discussion on quasi-helical stellarators that motivated this work.

This work has been carried out within the framework of the EUROfusion Consortium, funded by the European Union via the Euratom Research and Training Programme (Grant Agreement No 101052200 — EUROfusion). Views and opinions expressed are however those of the author(s) only and do not necessarily reflect those of the European Union or the European Commission. Neither the European Union nor the European Commission can be held responsible for them.

\section*{Data availability statement}
\hlchangedrev{The simulation data presented in this paper and the python scripts for the boundary surface computations in section~\ref{sec:quasr} are openly available at 
%OLD%\url{https://doi.org/10.5281/zenodo.13960848}.
\url{https://doi.org/10.5281/zenodo.14714598}.}
%\bibliographystyle{unsrt}
% % Note the spaces between the initials

%\bibliography{references}
\appendix
\section{\hlchangedrev{Behavior of the Jacobian of the map $\hmap$ with the \GFF{} }\label{app:jacobian}}
%When computing variations of the MHD energy \eqref{eq:energy} with respect to $X^1,X^2$, second derivatives of the map $\hmap$ are needed. 
\hlchangedrev{Given the map $\hmap$ in \eqref{eq:gff_h} using the \GFF{}, only first derivatives of $\X,\Nnew,\Bnew$, as already used in \eqref{eq:metrics_h}, will be necessary.
The Jacobian determinant is defined as
\begin{equation}
  \Jh \coloneqq \ddp{h}{q^3}\cdot\left(\ddp{h}{q_1}\times\ddp{h}{q_2} \right)=\ttilde\cdot(\Nnew\times \Bnew) \quad \text{with}\quad \ttilde:= \X'+q^1 \Nnew' +q^2 \Bnew'. \label{eq:jacobian_h}
\end{equation}
% \begin{equation}
% \begin{aligned}
%   \Jh^2 &\coloneqq\det(G) =  G_{11}G_{22}G_{33} + 2G_{12}G_{23}G_{31}  
%   -G_{13}^2 G_{22}  - G_{23}^2 G_{11} - G_{12}^2 G_{33} \\
%  & =|\Nnew|^2|\Bnew|^2|\ttilde|^2 + 2 (\Nnew\cdot \Bnew)(\Bnew\cdot \ttilde) (\Nnew\cdot \ttilde) \\& \qquad - (\Nnew\cdot \ttilde)^2|\Bnew|^2-(\Bnew\cdot \ttilde)^2|\Nnew|^2 - (\Nnew\cdot \Bnew)^2|\ttilde|^2 \\
%   & =|\Nnew\times \Bnew|^2|\ttilde|^2
%   + ((\Nnew\times \Bnew)\cdot(\Bnew\times \ttilde))(\Nnew\cdot \ttilde) 
%   +((\Bnew\times \Nnew)\cdot(\Nnew\times \ttilde))(\Bnew\cdot \ttilde)\,,
%   %& =(\Nnew\times \Bnew)\cdot\left[(\Nnew\times \Bnew)(\ttilde\cdot\ttilde)
%   %+ (\Bnew\times \ttilde)(\Nnew\cdot \ttilde) 
%   % -(\Nnew\times \ttilde))(\Bnew\cdot \ttilde)\right]\\  
% \end{aligned} \label{eq:jacobian_h}
% \end{equation}
%where we used the identity $(a\times b)\cdot(c\times d)= (a\cdot c)(b\cdot d)-(a\cdot d)(b\cdot c)$ holding for any $a,b,c,d\in \mathbb{R}^3$. 
The derivatives of the Jacobian determinant are conveniently computed from \eqref{eq:jacobian_h}, 
\begin{equation}
    \ddp{\Jh}{q^1}= \Nnew'\cdot(\Nnew\times \Bnew)\,,\quad  \ddp{\Jh}{q^2}= \Bnew'\cdot(\Nnew\times \Bnew). \label{eq:jacobian_h_dq}
\end{equation}
}%\hlchangedrev

\hlchangedrev{Note that these identities can be computed directly, using the \GFF{} interface from section~\ref{sec:gff_interface}.}

\begin{figure}[htbp!]
    \centering
    \includegraphics[trim=0 10 0 0,clip,width=0.98\textwidth]{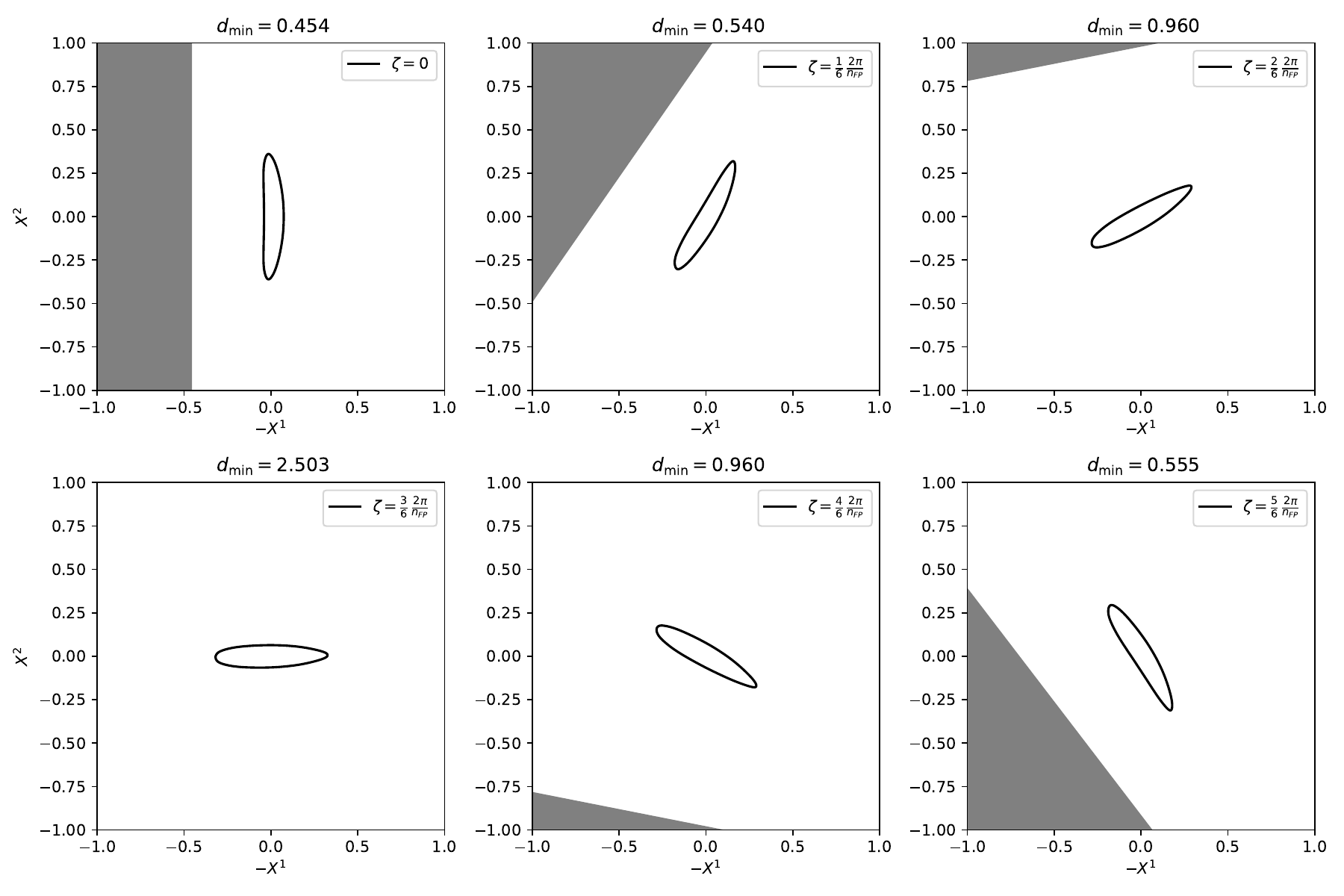}\\
    \caption{\hladdedrev{Cross-sections in the $-X^1,X^2$ coordinates of the \GFF{}, computed from the boundary surface presented in section~\ref{sec:quasr}. The region of negative Jacobian is shown in gray, at the minimum distance from the origin $d_\text{min}$.}}
    \label{fig:quasr_Jlimits}
\end{figure}

\hladdedrev{As the Jacobian $\Jh$ is linear in $q^1,q^2$, we can express it as
\begin{equation}
   \Jh(q^1,q^2,q^3) = \mathcal{J}_0 + q^1 \mathcal{J}_1 + q^2 \mathcal{J}_2, 
   %%%\left .\ddp{\Jh}{q^1}\right |_{0} q^1 +\left .\ddp{\Jh}{q^2}\right |_{0} q^2\,.
\end{equation}
where $\mathcal{J}_i$, for $i = 0,1,2$, depends only on $q^3 = \zeta$, with
\begin{equation}
\mathcal{J}_0 = \Xp\cdot(\Nnew\times \Bnew), \qquad \mathcal{J}_1 = \Nnew'\cdot(\Nnew\times \Bnew), \qquad \mathcal{J}_2 = \Bnew'\cdot(\Nnew\times \Bnew).
\end{equation}
In each cross-section ($\zeta$=const.), the point of minimum distance in $q^1,q^2$ where the Jacobian becomes zero is
\begin{equation}
   q^1_{\text{min}} = -\frac{\mathcal{J}_0 \mathcal{J}_1 }{ \mathcal{J}_1^2 + \mathcal{J}_2^2}, \quad
   q^2_{\text{min}} = -\frac{\mathcal{J}_0 \mathcal{J}_2 }{ \mathcal{J}_1^2 + \mathcal{J}_2^2}. 
   %%%q^1_\text{min} =\frac{-\left. \Jh \right |_{0}  \left.\ddp{\Jh}{q^1}\right |_{0} }{\left(\left. \ddp{\Jh}{q^1}\right |_{0} \right)^2+\left(\left. \ddp{\Jh}{q^2}\right |_{0} \right)^2} \,, \quad 
   %%%q^2_\text{min} = \frac{-\left. \Jh \right |_{0}  \left.\ddp{\Jh}{q^2}\right |_{0} }{\left(\left. \dd p{\Jh}{q^1}\right |_{0} \right)^2+\left(\left. \ddp{\Jh}{q^2}\right |_{0} \right)^2}\,.
\end{equation}
The distance from the origin is then 
\begin{equation}
    d_\text{min}=\sqrt{(q^1_\text{min})^2+(q^2_\text{min})^2} = \frac{ \mathcal{J}_0 }{\sqrt{ \mathcal{J}_1^2 + \mathcal{J}_2^2}},
    %%%%d_\text{min}=\sqrt{(q^1_\text{min})^2+(q^2_\text{min})^2}=\frac{\left. \Jh \right |_{0}}{\left(\left. \ddp{\Jh}{q^1}\right |_{0} \right)^2+\left(\left. \ddp{\Jh}{q^2}\right |_{0} \right)^2} \,,
\end{equation}
 and the line of zero Jacobian is then expressed as 
\begin{equation}
    \left(q^1_\text{min} -q^2_\text{min} t, \quad q^2_\text{min} +q^1_\text{min} t)\right),
\end{equation}
which limits the domain of positive Jacobian in $q^1,q^2$ to one side.
}%hladdedrev

\hladdedrev{Taking the example of the \GFF{} computed from the boundary surface in section~\ref{sec:quasr}, one can visualize the cross-sections along with the region of negative Jacobian, as shown in Figure~\ref{fig:quasr_Jlimits}. At $\zeta=0$ the distance is the smallest, but still allows for a much larger boundary surface, thus allowing to increase the aspect ratio while keeping the \GFF{} fixed. Also, it can be seen that the major axis of the cross-section tends to align with the line of zero Jacobian.
}%hladdedrev

\section{\hladdedrev{Rotated ellipse representation in 2D}\label{app:ellipse}}
\newcommand\rotangle{{\Gamma_e}}
\hladdedrev{
In this section, we relate the representation of a centered rotated ellipse in two dimensions to its Fourier coefficients.}
\begin{figure}[htbp!]
    \centering
    \includegraphics[width=0.4\textwidth]{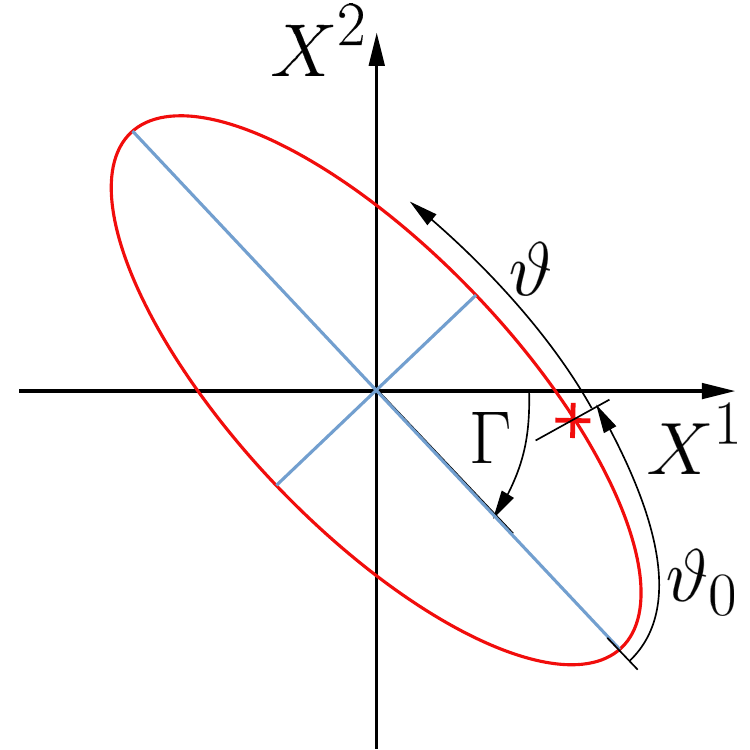}\\
    \caption{\hladdedrev{Sketch of the centered rotated ellipse in the 2D plane, with coordinates $X^1,X^2$, rotation angle $\rotangle$ and offset $\thet_0$}}\label{fig:rotated_ellipse}
    \end{figure}
 \hladdedrev{As shown in Figure~\ref{fig:rotated_ellipse}, the general form of a rotated ellipse centered at the origin of $X^1,X^2$ is  
\begin{equation}
    \begin{aligned}
    X^1(\thet) &= \quad \cos(\rotangle) a \cos(\thet+\thet_0)+ \sin(\rotangle) b \sin(\thet+\thet_0)\,, \\
     X^2(\thet)&= -\sin(\rotangle) a \cos(\thet+\thet_0)+ \cos(\rotangle) b \sin(\thet+\thet_0)\,,
    \end{aligned}\label{eq:rot_ell}
\end{equation}
with minor radius $a$, major radius $b$, rotation angle $\rotangle$ and shift $\thet_0$. Another equivalent form is using four coefficients $x^1_c,x^1_s,x^2_c,x^2_s$ and,
\begin{equation}
    \begin{aligned}
     X^1(\thet)&= x^1_c\cos(\thet)+ x^1_s\sin(\thet)\,, \\
     X^2(\thet)&= x^2_c\cos(\thet)+ x^2_s\sin(\thet)\,.
    \end{aligned}
\end{equation}
We write the coefficients in terms of $\epsp=(a+b)/2,\epsm=(a-b)/2,\rotangle,\thet_0$ as
\begin{equation}
    \begin{aligned}
        x^1_c&=\quad \epsp\cos(\rotangle-\thet_0)+\epsm\cos(\rotangle+\thet_0)\,, \\
        x^1_s&=\quad \epsp\sin(\rotangle-\thet_0)-\epsm\sin(\rotangle+\thet_0)\,,\\
        x^2_c&=-\epsp\sin(\rotangle-\thet_0)-\epsm\sin(\rotangle+\thet_0)\,,\\
        x^2_s&= \quad\epsp\cos(\rotangle-\thet_0)-\epsm\cos(\rotangle+\thet_0)\,,
    \end{aligned}
\end{equation}
and then express the parameters of \eqref{eq:rot_ell} in terms of the Fourier coefficients as
\begin{equation}
    a=\epsp+\epsm,\quad b=\epsp-\epsm\, \,,\quad  \varepsilon^{(\pm)}=\frac{1}{2}\sqrt{(x^1_c\pm x^2_s)^2+(x^1_s \mp x^2_c)^2}
 \,.
\end{equation}
and 
\begin{equation}
    \rotangle\pm\thet_0 = \arctan\left(\frac{x^1_s\pm x^2_c}{x^2_s \mp x^1_c} \right)\,. \label{eq:rot_ell_Gamma_pm_thet0}
\end{equation}
Note that the rotation function $\Gamma(\zeta)$ shown in Figure~\ref{fig:quasr_untwist} is computed from $(\thet_0-\rotangle)$.
 }%hladdedrev
\section*{References}
\bibliographystyle{iopart-num}
\bibliography{references}

\providecommand{\newblock}{}
\begin{thebibliography}{10}
\expandafter\ifx\csname url\endcsname\relax
  \def\url#1{{\tt #1}}\fi
\expandafter\ifx\csname urlprefix\endcsname\relax\def\urlprefix{URL }\fi
\providecommand{\eprint}[2][]{\url{#2}}
% Bibliography created with iopart-num v2.1
% /biblio/bibtex/contrib/iopart-num

\bibitem{landreman-sengupta-plunk}
Landreman M, Sengupta W and Plunk G~G 2019 {\em Journal of Plasma Physics\/}
  {\bf 85} 905850103

\bibitem{plunk_landreman_helander_2019}
Plunk G~G, Landreman M and Helander P 2019 {\em Journal of Plasma Physics\/}
  {\bf 85} 905850602

\bibitem{Giuliani_2024_quasr}
Giuliani A 2024 {\em Journal of Plasma Physics\/} {\bf 90} 905900303

\bibitem{bishop_frame}
Bishop R~L 1975 {\em The American Mathematical Monthly\/} {\bf 82} 246--251

\bibitem{yilmaz_new_bishop}
Yilmaz S and Turgut M 2010 {\em Journal of Mathematical Analysis and
  Applications\/} {\bf 371} 764--776 ISSN 0022-247X

\bibitem{rotation_minimizing_frame}
Wang W, J\"{u}ttler B, Zheng D and Liu Y 2008 {\em ACM Trans. Graph.\/} {\bf
  27} ISSN 0730-0301 \urlprefix\url{https://doi.org/10.1145/1330511.1330513}

\bibitem{hirshman_VMEC_1983}
Hirshman S~P and Whitson J~C 1983 {\em Physics of Fluids\/} {\bf 26} 3553--3568
  ISSN 0031-9171 \urlprefix\url{https://doi.org/10.1063/1.864116}

\bibitem{Hirshman1986_vmec_free}
Hirshman S, van RIJ W and Merkel P 1986 {\em Computer Physics Communications\/}
  {\bf 43} 143–155 ISSN 0010-4655
  \urlprefix\url{http://dx.doi.org/10.1016/0010-4655(86)90058-5}

\bibitem{Hirshman1991_vmec_precond}
Hirshman S and Betancourt O 1991 {\em Journal of Computational Physics\/} {\bf
  96} 99–109 ISSN 0021-9991
  \urlprefix\url{http://dx.doi.org/10.1016/0021-9991(91)90267-O}

\bibitem{Huang_2023}
Huang Y~M, Zhou Y, Loizu J, Hudson S and Bhattacharjee A 2023 {\em Plasma
  Physics and Controlled Fusion\/} {\bf 65} 129601

\bibitem{Dhaeseleer1991}
D’haeseleer W~D, Hitchon W~N~G, Callen J~D and Shohet J~L 1991 {\em Flux
  Coordinates and Magnetic Field Structure\/} (Springer Berlin Heidelberg) ISBN
  9783642755958 \urlprefix\url{http://dx.doi.org/10.1007/978-3-642-75595-8}

\bibitem{dudt_DESC_2020}
Dudt D~W and Kolemen E 2020 {\em Physics of Plasmas\/} {\bf 27} 102513 ISSN
  1070-664X \urlprefix\url{https://doi.org/10.1063/5.0020743}

\bibitem{gvec-2019}
Hindenlang F, Maj O, Strumberger E, Rampp M and Sonnendrücker E 2019 {GVEC}: a
  newly developed {3D} ideal {MHD} {G}alerkin variational equilibrium code
  annual Meeting of the Simons Collaboration on Hidden Symmetries and Fusion
  Energy

\bibitem{garren-boozer-1}
Garren D~A and Boozer A~H 1991 {\em Physics of Fluids B\/} {\bf 3} 2805--2821
  \urlprefix\url{http://scitation.aip.org/content/aip/journal/pofb/3/10/10.1063/1.859915}

\bibitem{garren-boozer-2}
Garren D~A and Boozer A~H 1991 {\em Physics of Fluids B\/} {\bf 3} 2822--2834
  \urlprefix\url{http://scitation.aip.org/content/aip/journal/pofb/3/10/10.1063/1.859916}

\bibitem{landreman-sengupta}
Landreman M and Sengupta W 2018 {\em Journal of Plasma Physics\/} {\bf 84}
  905840616

\bibitem{Landreman_Sengupta_2019}
Landreman M and Sengupta W 2019 {\em Journal of Plasma Physics\/} {\bf 85}
  815850601

\bibitem{camacho-mata_plunk_2022}
Camacho~Mata K, Plunk G~G and Jorge R 2022 {\em Journal of Plasma Physics\/}
  {\bf 88} 905880503

\bibitem{plunk2024-QI}
Plunk G~G {\em et~al.\/} 2024 A geometric approach to constructing
  quasi-isodynamic fields {\em in preparation}

\bibitem{simsopt-joss}
Landreman M, Medasani B, Wechsung F, Giuliani A, Jorge R and Zhu C 2021 {\em
  Journal of Open Source Software\/} {\bf 6} 3525
  \urlprefix\url{https://doi.org/10.21105/joss.03525}

\bibitem{tecchiolli_constructing_2024}
Tecchiolli Z, Hudson S, Loizu J, Köberl R, Hindenlang F and De~Lucca B 2024
  {\em submitted to Journal of Plasma Physics\/} ArXiv:2405.08173
  \urlprefix\url{http://arxiv.org/abs/2405.08173}

\bibitem{plunk2024figure8}
Plunk G~G, Drevlak M, Rodriguez E, Babin R, Goodman A and Hindenlang F 2024
  Back to the figure-8 stellarator (\textit{Preprint} \eprint{2411.16411})
  \urlprefix\url{https://arxiv.org/abs/2411.16411}

\end{thebibliography}
\end{document}